\documentclass[11pt]{article}
\usepackage[a4paper, total={6.3in, 9.7in}]{geometry}
\usepackage[pdfusetitle]{hyperref}
\usepackage{setspace} 
\usepackage{authblk}
\usepackage{graphicx}
\usepackage[dvipsnames,svgnames,x11names]{xcolor}
\usepackage{doi}
\usepackage{cite}
\usepackage{booktabs}
\usepackage{amsmath,mathtools,amssymb}

\usepackage{etoolbox}
\usepackage{multirow}

\makeatletter
\patchcmd{\env@cases}{1.2}{0.72}{}{}
\makeatother

\hypersetup{
    colorlinks,
    linkcolor={red!50!black},
    citecolor={blue!50!black},
    urlcolor={blue!80!black}
}

\title{\Large\bf
LeStrat-Net: Lebesgue style stratification for Monte Carlo simulations powered by machine learning}
\author[1]{Kayoung Ban\footnote{\href{mailto:kban@kias.re.kr}{kban@kias.re.kr}}}
\author[1,2,3]{Myeonghun Park\footnote{\href{mailto:parc.seoultech@seoultech.ac.kr}{parc.seoultech@seoultech.ac.kr}}}
\author[4]{Raymundo Ramos\footnote{\href{mailto:raramos@kias.re.kr}{raramos@kias.re.kr}}}
\affil[1]{\textit{School of Physics, Korea Institute for Advanced Study, Seoul 02455, KOREA}}
\affil[2]{\textit{Institute of Convergence Fundamental Studies, Seoultech, Seoul 01811, KOREA}}
\affil[3]{\textit{School of Natural Sciences, Seoultech,  Seoul 01811, KOREA}}
\affil[4]{\textit{Quantum Universe Center, Korea Institute for Advanced Study, Seoul 02455, KOREA}}

\date{}

\begin{document}

\maketitle

\begin{abstract}
    We develop a machine learning algorithm to turn around stratification in Monte Carlo sampling.
    We use a different way to divide the domain space of the integrand,
    based on the height of the function being sampled,
    similar to what is done in Lebesgue integration.
    This means that isocontours of the function
    define regions that can have any shape
    depending on the behavior of the function.
    We take advantage of the capacity of neural networks
    to learn complicated functions
    in order to predict these complicated divisions
    and preclassify large samples of the domain space.
    From this preclassification we can select the required number of points to perform
    a number of tasks such as variance reduction, integration and even event selection.
    The network ultimately defines the regions with what it learned and is also used to calculate
    the multi-dimensional volume of each region.
\end{abstract}

\newpage

\section{Introduction}
\label{sec:intro}

In high energy physics, a crucial operation involves computing scattering amplitudes and integrating them across the phase space. 
Due to the complex multidimensional character of these amplitudes, which are frequently encountered in detector simulations, numerical strategies are necessary for integration. 
This method involves iteratively evaluating the amplitudes at discrete points throughout the phase space. 

Recent advances have seen the integration of machine learning (ML) techniques to enhance the efficiency and accuracy of event generation and integration processes. Boosted Decision Trees (BDT) and artificial neural networks have been employed as event generators, offering significant improvements in performance and precision~\cite{Bendavid:2017zhk, Klimek:2018mza, Chen:2020nfb, Otten:2019hhl, Bellagente:2019uyp, Bothmann:2020ywa, Danziger:2021eeg, Janssen:2023ahv, Bothmann:2023siu}.
In addition to event generation, ML techniques have also been applied to the training with amplitude values. This approach leverages the predictive power of neural networks to estimate physical amplitudes directly from data ~\cite{Bishara:2019iwh, Maitre:2021uaa}.
Furthermore, normalizing flows and invertible neural networks have been introduced for phase space integration~\cite{Gao:2020zvv, Heimel:2022wyj, Ernst:2023qvn, Buss:2024orz}.

In this work, we propose an approach to stratified sampling that leverages ML techniques, particularly neural networks, to create regions based on the size of the integrand.
Traditional stratified sampling methods typically divide the domain based on coordinate components or regions with large variances.
With our method, we aim to create regions limited by isocontours of the integrand, analogous to Lebesgue integration.
Intuitively, Lebesgue integration divides the integrand range in intervals that are
then translated into partitions of the domain space~\cite{rudin1976principles}.
In general this will result in regions of arbitrary shape or even composed by multiple subregions.
Modern ML techniques facilitate learning these complicated divisions
and can provide fast and accurate approximations of the boundaries,
enabling efficient and effective determination of regions in a large data set.
To achieve this, we employ neural networks to estimate the classification of points into their corresponding regions.
This involves training neural networks to predict for every point either a subregion index or the position with respect to the isocontours.
After we have trained the network, we can accurately partition the integration domain and optimize the allocation of sample points to minimize variance.
We enhance the accuracy of the predictions
by training the neural network with the method of Ref.~\cite{Hammad:2022wpq}.
In said reference, a neural network is iteratively trained
on some complicated function,
and used in every iteration to suggest more points to explore.

Our paper is structured as follows:
In Sec.~\ref{sec:MCIntegration} we summarize details about Monte Carlo integration
that are relevant to our work.
In Sec.~\ref{sec:MCfsize} we describe the important details
of dividing the domain space using isocontours of the integrand size
including what this implies for variance.
Section~\ref{sec:NNSurvey} contains the full details of how we propose
to use neural networks to learn to classify regions of the domain space,
expand our knowledge of our calculation and creating more regions in needed.
Here we also describe the reasoning of our approach.
In Sec.~\ref{sec:examples} we present examples
of how the neural networks can be applied
to gather points and perform integration,
including information on variance reduction and required sample size.
We also include a proposal of how to obtain unweighted events
from the preclassification made by the network.
Finally, in Sec.~\ref{sec:conclusion} we conclude by summarizing the main details of our work and
comment on possible extensions.

\section{Monte Carlo integration}
\label{sec:MCIntegration}

Several techniques exist to perform numerical integration of arbitrary functions.
In low dimensions, some deterministic techniques,
like numerical quadrature,
are quite performant and provide precise results
with rapidly decreasing errors.
However, such techniques
tend to require a number of points
that grows exponentially with the number of dimensions.
The favored technique for high dimensional integration is Monte Carlo methods,
where the error can be reduced by the squared root of the number of samples,
irrespective of the number of dimensions.

The basic idea behind Monte Carlo integration
is the realization that the value of the integral of a function $f:\mathbb{R}^d \to \mathbb{R}$
over a $d$-dimensional section of space $\Phi$ with volume $V_\Phi$ can be expressed as
\begin{equation}
    I[f] = \int_\Phi dx\, f(x) = V_\Phi \langle f \rangle_\Phi, \quad\text{with}\quad V_\Phi = \int_\Phi dx\,.
\end{equation}
Here $x$ represents a random $d$-dimensional vector $x\in\Phi$ of the domain,
uniformly distributed,
and $\langle f \rangle_\Phi$ stands for the expected value of $f$
in the section of space $\Phi$
where the integral is being performed.
Therefore, an estimate of $I$ can be obtained
by sampling $N$ vectors $x_n$ from $\Phi$,
evaluating the corresponding $f(x_n)$
and calculating the average times volume of integration
\begin{equation}
    \label{eq:MCEstimate}
    E(I) = \frac{V_\Phi}{N} \sum_{n=1} ^N f(x_n);\quad\text{with}\quad x_n\in\Phi\,,
\end{equation}
where we have used $E(I)$ to indicate estimate of $I$.
To evaluate the precision of the estimate in Eq.~\eqref{eq:MCEstimate},
we can use the central limit theorem to calculate
the variance of the estimate, $\sigma^2(E(I))$,
from the variance of $f$
\begin{equation}
    \label{eq:MCVariance}
    \sigma^2(E(I)) = \frac{V_\Phi^2\sigma^2(f)}{N} = \frac{V_\Phi^2}{N} \left[ 
        \frac{1}{V_\Phi} \int_\Phi dx\, \left(f(x) - \langle f \rangle \right)^2
    \right].
\end{equation}
Again, we can use a sample of $N$ vectors of the domain space to estimate this integral
\begin{align}
    \label{eq:MCVarianceEst}
    \sigma^2(E(I)) \approx \frac{V_\Phi^2}{N}\left[
        \frac{1}{N}\sum_{n=1}^N f^2(x) - \frac{1}{N^2}\left( \sum_{n=1}^N f(x_n)\right)^2
    \right]\,.
\end{align}
From Eq.~\eqref{eq:MCVariance} it is clear that the error of our estimate,
$\sigma(E(I))$ (the square root of Eq~\eqref{eq:MCVarianceEst}),
shrinks by a factor of $1/\sqrt{N}$ regardless of the dimensionality.

\subsection{Variance reduction with stratified sampling}

As pointed out above, the error of the Monte Carlo estimation of the integral
is reduced by $\sqrt{N}$
independently of the dimensionality.
However, such a reduction in error is slow
and may require enormous samples
for functions with large variation.

Many techniques exist to reduce the variance in Monte Carlo integration.
In this work we are interested in a particular technique known as \emph{stratified sampling}.
This technique takes advantage of the additive interval property of integrals,
that can be expressed as
\begin{equation}
    I[f] = \int_\Phi dx\, f(x) = \sum_{j=1}^M \int_{\Phi_j} dx\, f(x)\, = \sum_{j=1}^M \int_{\Phi_j} I_j\,,
    \quad\quad
    \Phi = \sum_{j=1}^M \Phi_j
\end{equation}
where every $\Phi_j$ is disjoint with all the others.
In this case, we can calculate partial estimations
that can be added up to obtain the total estimation
\begin{equation}
    E(I_j) = \frac{V_{\Phi_j}}{N_j} \sum_{n=1}^{N_j} f(x_n\in\Phi_j)\,,
    \quad
    E(I) = \sum_{j=1}^M E(I_j)\,,
    \quad
    V_{\Phi_j} = \int_{\Phi_j} dx\,.
\end{equation}
In the same way,
the variance of the estimation is now given by
\begin{equation}
    \label{eq:SSVariance}
    \sigma_\text{ss}^2(E(I)) = \sum_{j=1}^M \frac{V_{\Phi_j}^2}{N_j} \sigma^2_{\Phi_j}(f)
\end{equation}
that now depends on the variances of each subregion
\begin{equation}
    \sigma_{\Phi_j}^2(f) = \frac{1}{V_{\Phi_j}} \int_{\Phi_j} dx\, \left(
        f(x) - \langle f \rangle_{\phi_j}
    \right)^2\,,
    \quad
    \langle f \rangle_{\Phi_j} = \frac{1}{V_{\Phi_j}} \int_{\Phi_j} f(x)\,.
\end{equation}
As in the previous subsection,
we can estimate these variances with random samples from each subregion
\begin{equation}
    \label{eq:regVariance}
    \sigma^2_{\Phi_j}(f) \approx
        \frac{1}{N_j} \sum_{n = 1}^{N_j} f^2(x_n\in \Phi_j)
        - \frac{1}{N_j^2} \left( \sum_{n = 1}^{N_j} f(x_n\in \Phi_j) \right)^2\,.
\end{equation}
Error reduction works the same as before for each subregion,
being reduced by a factor of $1/\sqrt{N_j}$.
This provides a simple general rule for reduction of variance in Eq.~\eqref{eq:SSVariance},
since we just need to take the number of points in each region, $N_j$,
to be proportional to the corresponding $\sigma_{\Phi_j}(f)$.

\section{Stratified Monte Carlo based on integrand size}
\label{sec:MCfsize}

The usual approach to stratified sampling is to create regions
based on divisions of the coordinate components of the space $\Phi$,
estimate the variance in each region,
and determine the number of points required.
More advanced techniques can keep this process by, e.g.,
creating more subdivisions in regions with larger variance.
In this work, we want to present an alternative approach to creating subregions
of the integration space,
namely, subregions based on the size of the integrand.
While such an approach requires an accurate estimation of the integrand size
in the whole space,
modern ML techniques simplify the task of creating
functions that are capable of achieving this accurate estimation
while being fast to evaluate.

First, start with a function $f:\mathbb{R}^d \to \mathbb{R}$
that we want to integrate in some $d$-dimensional space $\Phi$.
Then, we proceed to divide our $d$-dimensional integrand
into $n$ sections based on integrand value.
The $n$ sections are limited by $n + 1$ isocontours with values $l_j$ and $j=0,1,2,\ldots,n$.
Each of these sections defines regions of the domain according to
\begin{equation}
    \label{eq:domaindivisions}
    \Phi_j = \left\{x \mid l_{j-1} < f(x) \leq l_j \right\}.
\end{equation}
Considering that,
in most cases,
$l_0$ and $l_n$ can be safely assumed to be $-\infty$ and $\infty$,
respectively, sometimes we will only need to determine $n - 1$ isocontours.
In general, each $\Phi_j$ can be composed of many disconnected subregions.
The integral considering this division is given by
\begin{equation}
    \label{eq:integration}
    I_\Phi [f] = \int_\Phi \mathrm{d}^d x \, f(x) = \sum_{j=1}^n \int_{\Phi_j} \mathrm{d}^d x\, f(x) = \sum_{j=1}^n V_{\Phi_j} \langle f\rangle_{\Phi_j}
\end{equation}
where $V_{\Phi_j} = \int_{\Phi_j} \mathrm{d}^d x$
and $\langle f\rangle_{\Phi_j} = (\int_{\Phi_j} \mathrm{d}^d x f(x))/(\int_{\Phi_j} \mathrm{d}^d x)$
were used.
The last expression in Eq.~\eqref{eq:integration} can easily be adapted for Monte Carlo integration.
For an uniformly distributed set of $N$ random points in $\Phi$ space
the $d$-dimensional volume of a subspace or region $\Phi_j$ can be estimated as
\begin{equation}
    \label{eq:volnphi}
    E(V_{\Phi_j}) \approx \frac{N_j}{N} V_{\Phi}
\end{equation}
where $N_j$ is the count of points in $\Phi_j$ from the total $N$ points.
The sampling does not need to happen in the full space $\Phi$
but it is enough that the sampling is done
in a subspace $\Phi_\text{sam}$
such that $\Phi_j \subset \Phi_\text{sam}$.
In that case we need to replace $V_\Phi$ in Eq.~\eqref{eq:volnphi}
with the volume of $\Phi_\text{sam}$.
The average is estimated by
\begin{equation}
    \label{eq:intnphi}
    E(\langle f \rangle_{\Phi_j}) = \frac{1}{N_j}\sum_{i = 1}^{N_j} f(x_i).
\end{equation}
The set of points used for integration in Eq.~\eqref{eq:volnphi}
does not need to be the same set of points used in Eq.~\eqref{eq:intnphi}.
Obviously, the average $\langle f \rangle_{\Phi_j}$
has a value between $l_{j - 1}$ and $l_j$.
Once the volume $V_{\Phi_j}$ of each section is accurately determined,
the contribution of each section is bounded by $V_{\Phi_j} l_{j - 1}$
and $V_{\Phi_j} l_j$.

\subsection{Creating divisions of the domain}
\label{sec:dividedomain}

Considering that we may use as many divisions as desired and that the distancing between divisions can follow any rule, there are infinite ways to create divisions of the domain.
Among the simplest rules, there is the possibility of considering a single division and also divisions with equal spacing in values of the domain.
Let's call $\Delta f_j \equiv l_j - l_{j - 1}$ to the spacing between two limits $l_{j - 1}$ and $l_{j}$.
A division with equal spacing would have the property $\Delta f_j = \Delta f_k$
for any section with index $j$ and $k$.
Another simple example, for cases when we want to pay attention to lower values of $f(x)$, is to use equal spacing on values of $\log(f)$.
This requires that the first limit follows $l_0 > 0$ to avoid the divergence at $\log(0)$.
In this case, the limits would have the property $l_j/l_{j-1} = l_k/l_{k - 1}$.
We could create more complicated rules
that take account of some of the properties of length,
the domain space, the distribution of $f(x)$,
and practically any other characteristics that may affect the integration and the sampling of the space.
One example is the creation of subdivisions based on equal distancing of
the cumulative length, $C_L$.
The cumulative length can be defined as the length that has accumulated
in going from some initial value $f_0$ to some value $f_L$.
Using again $V$ to represent the length,
and $V([a, b])$ to represent the length
for some interval where $a < f(x) \leq b$,
we define
\begin{equation}
    \label{eq:cumulv}
    C_L(f_L) \equiv V([f_0, f_L])\,.
\end{equation}
It may not be immediately obvious from the equation above,
but equal spacing on $C_L(f_L)$
has the property
\begin{align}
    C_L(l_{j}) - C_L(l_{j - 1}) & = C_L(l_{k}) - C_L(l_{k-1}) \\
    V([f_0, l_{j}]) - V([f_0, l_{j - 1}]) & = V([f_0, l_{k}]) - V([f_0, l_{k - 1}]) \\
    V([l_{j-1}, l_{j}]) & = V([l_{k - 1}, l_{k}])
\end{align}
which means that each section has the same length as the others.
In the sequence of equations above we used
$V([a, c]) = V([a, b]) + V([b, c])$, with $a < b \leq c$.

One more way to subdivide the domain,
the one we will use in some examples,
is to divide in terms of the contribution to the integral.
Similarly as with the length,
we define a cumulative contribution to the integral value, $C_I$
\begin{equation}
    \label{eq:cumuli}
    C_I(f_I) = \langle f\rangle_{[f_0, f_I]} V([f_0, f_I]) \,,
\end{equation}
where $\langle f\rangle_{[f_0, f_I]}$ represents the average value of $f(x)$
considering only the points where $f_0 < f(x) \leq f_I$.
In this case, subdivisions with equal distance in $C_I(f_I)$ represent
sections with equal contribution to the integral.
This conclusion follows from $\langle f\rangle_{[f_0, f_I]} = \int_{[f_0, f_I]} f \mathrm{d}x /\int_{[f_0, f_I]} \mathrm{d}x$
and $V([f_0, f_I]) = \int_{[f_0, f_I]} \mathrm{d}x$.

Another interesting example of how to subdivide the integration domain
is to consider the variance of the created regions.
This is particularly important for controlling the variance of subregions
during Monte Carlo integration.
From Eq.~\eqref{eq:SSVariance} we know that each region contributes $V_{\Phi_j}^2 \sigma_{\Phi_j}^2/N_j$.
We are interested in the contribution to the total variance
without dependence on the number of points,
given by $V_{\Phi_j}^2 \sigma_{\Phi_j}^2$
or its squared root.
Following the notation from this section,
let us represent this number as $V([l_{j-1},l_{j}])\sigma([l_{j-1},l_{j}])$
for the region where $l_{j-1} < f(x) \leq l_{j}$.
A convenient way to decide a value for this number
is to set it to a fraction of the value of the final integral $I[f]$,
such that the divisions themselves represent small contributions to the total variance.
If the same value is used for all regions,
that means that a similar number of points needs to be sampled from each region
to optimize the reduction of the variance.
In practice, the value of $I[f]$ is not known,
so a rough estimate can be used
to decide a value for $V([l_{j-1},l_{j}])\sigma([l_{j-1},l_{j}])$.

\subsection{Correction to variance}
\label{sec:fullvariance}

The variance obtained using Eqs.~\eqref{eq:SSVariance} and~\eqref{eq:regVariance}
considers only the contribution from estimating $\langle f \rangle_{\Phi_j}$
when the value of $V_{\Phi_j}$ is known.
However, for the idea expressed in the beginning of this section,
volumes of regions are also estimated using Eq.~\eqref{eq:volnphi}.
This means that we have another contribution
to the total variance of the integral estimated value.
For convenience, let us designate as $E(I_j)$ 
the estimated value of integral in region $\Phi_j$,
$E(V_{\Phi_j})$ the estimation of volume obtained using Eq.~\eqref{eq:volnphi}
and $E(\langle f \rangle_{\Phi_j})$ the estimation of the average of weights inside region $\Phi_j$.
Using this terminology and Eqs.~\eqref{eq:integration} to~\eqref{eq:intnphi},
the estimation of the integral in a single region is
\begin{equation}
    E(I_j) = E(V_{\Phi_j}) E(\langle f \rangle_{\Phi_j})\,.
\end{equation}
To calculate $E(V_{\Phi_j})$
numbers of points are used and their relationship
to the weights function $f(x)$ is only the limits $l_j$
of Eq.~\eqref{eq:domaindivisions},
that are also used to limit the values $f(x)$ to subspace $\Phi_j$.
From this we establish that the calculation of $E(V_j)$
is totally independent from the calculation of $E(\langle f \rangle_{\Phi_j})$.
In this case, the variance of $E(I_j)$ is composed of three terms
\begin{align}
    \label{eq:fullvar}
    \sigma^2(E(I_j)) & = V^2_{\Phi_j} \sigma^2(\langle f \rangle_{\Phi_j})
    + \langle f \rangle^2_{\Phi_j} \sigma^2(V_{\Phi_j})
    + \sigma^2(V_{\Phi_j})\sigma^2(\langle f \rangle_{\Phi_j})\,.
\end{align}
Obviously,
most of the times we will not have access
to the actual values of $V_{\Phi_j}$
and $\langle f \rangle_{\Phi_j}$.
In that case we can use estimates in Eq.~\eqref{eq:fullvar} as
\begin{align}
    \sigma^2(E(I_j))
    \label{eq:fullvarapprox}
    \approx E^2(V_{\Phi_j}) \sigma^2(\langle f \rangle_{\Phi_j})
    + E^2(\langle f \rangle_{\Phi_j}) \sigma^2(V_{\Phi_j})
    + \sigma^2(V_{\Phi_j})\sigma^2(\langle f \rangle_{\Phi_j})\,,
\end{align}
which could give a \emph{good enough} estimate of variance when
$E(V_{\Phi_j})$ and $E(\langle f \rangle_{\Phi_j})$
have been estimated down to a \emph{good enough} accuracy.
\emph{Good enough} here depends on the requirements of the problem being solved.
Note that the first term in Eqs.~\eqref{eq:fullvar} and~\eqref{eq:fullvarapprox}
corresponds to the typical Monte Carlo variance
when volumes are known, equivalent to the individual terms in Eq.~\eqref{eq:SSVariance}.
For Eq.~\eqref{eq:fullvarapprox}, volume values are replaced by estimations.
From Eq.~\eqref{eq:SSVariance} we can recognize $\sigma^2(\langle f \rangle_{\Phi_j}) = \sigma_{\Phi_j}^2(f)/N_j$.
To finish calculating Eq.~\eqref{eq:fullvarapprox}
we still need to give $\sigma^2(V_{\Phi_j})$.
If we sample $n_\text{test}$ points in an uniform distribution
in a space $\Phi_\text{sam}$ such that $\Phi_j \subset \Phi_\text{sam}$,
and $n_\text{in}$ of those points belong in $\Phi_j$,
as pointed out above, we can estimate the volume of $\Phi_j$ using
$V_\text{sam} n_\text{in}/n_\text{test}$,
where $V_\text{sam}$ is the volume of $\Phi_\text{sam}$.
This is equivalent to having $n_\text{in}$ points with
weights $w_V = V_\text{sam}$ and $n_\text{test} - n_\text{in}$ points
with weight $w_V = 0$.
Using this analogy, the variance of our estimation of volumes is given by
\begin{equation}
    \label{eq:volVariance}
    \sigma^2(V_{\Phi_j}) = \frac{\sigma^2(w_V)}{n_\text{test}} = \frac{V_\text{sam}^2}{n_\text{test}}\left(
        \frac{n_\text{in}}{n_\text{test}} - \frac{n^2_\text{in}}{n^2_\text{test}}
        \right)\,,
\end{equation}
which, again, decreases as $1/n_\text{test}$.
Note that to calculate $E(V_{\Phi_j})$ and $\sigma^2(V_{\Phi_j})$
we only need the volume of the sampled region, $V_\text{sam}$,
the number of points in the target region, $n_\text{in}$,
and the number of tested points, $n_\text{test}$.
Considering that it is standard to target $\sigma^2(\cdot)/E^2(\cdot) < 1$,
in Eqs.~\eqref{eq:fullvar} and~\eqref{eq:fullvarapprox}
we can expect the third term
to become negligible much faster than the other two.
The first and second terms on the total variance
are reduced by increasing the number of points
used in the estimations $E(\langle f \rangle_{\Phi_j})^2$ and $E(V_{\Phi_j})^2$,
respectively.
Such estimations do not need to use the same points
and can be reduced independently.
Here is where the advantage of dividing as described in Sec.~\ref{sec:MCfsize}
starts to become apparent.
If we reduce $\sigma^2(\langle f \rangle_{\Phi_j})$
by choosing regions that limit the range of $f(x)$,
we are effectively moving the difficult job to reducing $\sigma^2(V_{\Phi_j})$.
To really benefit from this whole discussion,
first we need to develop a process
that allows the precise calculation of $V_{\Phi_j}$ and,
thus, the reduction of its variance
by calculating as few values of $f(x)$ as possible.
In what follows we describe such process,
with a neural network at the center,
to take on the job of estimating the 
regions and their volumes.
In what follows we will describe a process
where a neural network takes the job of estimating
these regions and their volumes
and how this improves cases

\section{Neural networks for surveying and dividing the domain}
\label{sec:NNSurvey}

To determine regions as indicated in Eq.~\eqref{eq:domaindivisions},
we would need to run all the points that we wish to classify
through the calculation of the function $f$.
For time-consuming calculations and small regions of interest
this may represent a challenge
in terms of available time and computational resources.
It is here where it could be advantageous to have another function
that could estimate this classification
in a fraction of the time taken for the full calculation.
Such a function could accelerate the division process while also
help select only points deemed important for our calculation.
Our proposal is to use neural networks to estimate the regions
by using them to replace $f$ in Eq.~\eqref{eq:domaindivisions}:
\begin{equation}
    \label{eq:domaindivisionsml}
    \hat{\Phi}_j = \left\{x \mid \mathcal{N}\left(x; l_n, m_k\right) = j \right\}\,,
\end{equation}
where $\mathcal{N}\left(x; \left\{l_n\right\}, \left\{m_k\right\}\right)$ represents
a function that uses one or more neural networks that depend on the limits of $f$ given by $l_n$ and
the trainable parameters $m_k$.
In what follows, we will describe an implementation of this idea
to illustrate how neural networks can be used in the determination
of regions of importance during Monte Carlo sampling.

\subsection{Neural network output as indices of regions}
\label{sec:nnoutputindices}

The numerical output of a neural network
is determined by two characteristics of the output layer:
activation function and node configuration.
Considering that we want to adjust a series of indices denoting regions,
the simplest options for activation functions
include a linear function for regression or
\emph{softmax} for \emph{one-hot-encoded} indices.
In these two cases,
the conversion to integer indices
would be as simple as rounding the result for the linear case,
or applying an \emph{argmax} operation in the case of \emph{softmax}.
There are less trivial choices that could, however,
provide different opportunities in processing the output of the network
to improve accuracy during the fitting process.

\begin{figure}[tb]
    \includegraphics[width=\textwidth]{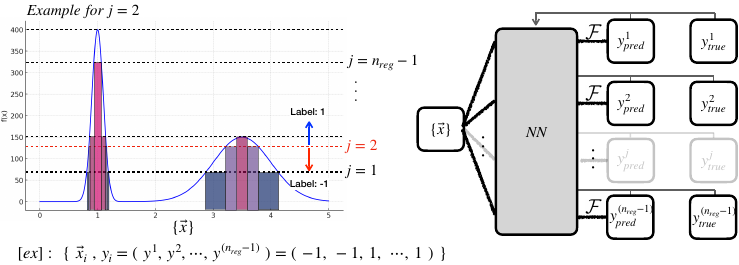}
    \caption{\label{fig:NNprocess}%
        The main purpose of the neural network is to help us classify the points
        into their corresponding regions. Here, $\mathcal{F}$ is the activation function.
        In the multilabel approach, for the point \( \{\vec{x}\} \), if \( f(\vec{x}) \) lies above the isocontour assigned to a particular value of \( j \), it is assigned a positive label, and if it is below, it receives a negative label (\emph{tanh} activation function). If a \emph{sigmoid} activation function is chosen, a label of zero is assigned instead of a negative label.
    }
\end{figure}

One example is to allow for a multilabel approach,
where the output of the networks is a vector
with each component the output of an individual activation function,
such as \emph{sigmoid} or \emph{tanh}.
Instead of attempting to use the combination of outputs
through \emph{argmax}, like it is done with \emph{softmax},
we could use each output component
to keep track of individual divisions
of the weight function.
This approach is described graphically in Fig.~\ref{fig:NNprocess}.
In this case,
the output of the neural network would have
$n_\text{reg} - 1$ components,
one for each limit, excluding absolute maximum and minimum.
Since each output component corresponds to labels of individual limits
we can identify this approach as \emph{multilabel}.
The conversion of network output
to predicted index of region
becomes slightly more involved in this case.
Consider an activation function, $\mathcal{F}$, with 
$\mathcal{F}_{min} < \mathcal{F} < \mathcal{F}_{max}$.
Also consider a network with multiple outputs using $\mathcal{F}$
as activation function for the output layer.
For a point $x$ the network returns the vector $\hat{\mathbf{Y}}(x)$,
with each component bounded as described above.
For convenience, for general $\mathcal{F}_{min}$ and $\mathcal{F}_{max}$, we can shift and rescale $\hat{\mathbf{Y}}(x)$
to limit it to the range (0,1):
\begin{equation}
    \label{eq:actprob}
    \hat{\mathbf{Y}}_{0,1} (x) = \frac{\hat{\mathbf{Y}}(x) - \mathcal{F}_\text{min} \times \mathbf{1}}{\mathcal{F}_\text{max} - \mathcal{F}_\text{min}}\,,
\end{equation}
where $\mathbf{1}$ is a vector full of 1 values and the same number of components
as $\hat{\mathbf{Y}}$.
With this adjustment,
we can easily obtain a predicted region index using
\begin{equation}
    \label{eq:actprobreg}
    \mathcal{N}(x) = r_0 + \sum_{j = 1}^{n_\text{reg} - 1} \left\lfloor 
    \left[\hat{\mathbf{Y}}_{0,1} (x)\right]_j + \frac{1}{2}
    \right\rfloor\,,
\end{equation}
where $\lfloor\cdot\rfloor$ indicates the floor function and
$r_0$ is the start of the indexing of the regions.
This equation represents rounding and summing
each element of $\hat{\mathbf{Y}}_{0,1} (x)$.

Here we want to point out two interesting differences between the \emph{softmax}
and \emph{multilabel} approaches.
To ease the discussion,
let us label this differences as
\emph{hierarchy of predictions}
and \emph{non-sense regions}.

\textbf{Hierarchy of predictions.}
Considering the information used during training,
in the \emph{softmax} case we employ vectors with a single
non-zero component in different positions to label different regions.
In this case, information about hierarchy between regions
(the difference in size between a region with large weights and another region
with smaller weights)
is left out of training when using categorical cross-entropy.
In the case of the \emph{multilabel} approach,
information about hierarchy of different regions
is added by including one label for each level for every point.
For a given point $x$, the training vector $\mathbf{Y}(x)$
has all components equal to 1 for the levels below the weight $f(x)$
and equal to 0 for the levels above.
This adds a sort of distance between vectors,
where regions higher in weight space
have larger amounts of 1s in $\mathbf{Y}(x)$.
Thus, it is difficult
to have jumps between two vectors
indicating distant regions, e.g.,
predict an all-0 vector
in the vicinity of points trained with all-1 vectors.

\textbf{Non-sense regions.}
Another interesting detail
comes from the fact that the multilabel approach
has many more possible combinations for $\mathbf{Y}(x)$
than actually used.
All valid outputs have, at most, two blocks,
a block of 1s followed by a block of 0s, e.g.,
(0, 0, 0), (1, 0, 0), (1, 1, 0) and (1, 1, 1) for 4 regions.
However, many more combinations are possible
that have no interpretation as indices of regions.
This leads to nonsensical regions
that can appear due to erroneous predictions
for uncertain regions.
Fortunately, such erroneous predictions can be identified and,
if not removed by typical loss functions,
reduced via extra contributions in a custom loss function.

\subsection{Survey to determine the limits}
\label{sec:survey}

\subsubsection{Single division}
\label{sec:survey_single}

Let us begin by describing the simplest unit in the process:
the determination of a single division between two regions of the domain.
First, assume that we have a set of points $x\in\Phi$,
their $f(x)$ values and
a function $\mathcal{C}(f_1, f_2)$ that describes some property
of the points $x$ such that $f_1 < f(x) \leq f_2$.
This function $\mathcal{C}$ will be matched to some criterion that sets
a value $f = l_1$ that divides the set $x\in\Phi$ in two.
The outline of the process is as follows:
\begin{enumerate}
    \item
        Calculate the value of $\mathcal{C}(f_0, f_\text{test})$ from an initial value $f_0$ to a test value $f_\text{test}$.
    \item
        If $f_0 < f_\text{test}$ ($f_0 > f_\text{test}$) keep increasing (decreasing) $f_\text{test}$
        until the condition $\mathcal{C}(f_0, f_\text{test}) = \mathcal{C}_1$ is met,
        where $\mathcal{C}_1$ is a number decided beforehand depending on the criterion
        (see Sec.~\ref{sec:dividedomain}).
        When the condition is met we set $l_1 = f_\text{test}$, i.e., $\mathcal{C}(f_0, l_1) = \mathcal{C}_1$.
    \item
        With the values $f(x)$ and the division $l_1$
        we can create labels $r(x)$ for each point, say
        $r(x) = 0$ if $f(x) \leq l_1$ and 1 for $f(x) > l_1$.
    \item
        Use $x$ and the labels $r(x)$ to train a neural network.
\end{enumerate}
At the end of this process we will have a model
that returns a number, $\hat{Y}_1(x)$, that represent the confidence that a given point belongs to $f(x) > l_1$,
and, conversely a $1 - \hat{Y}_1(x)$ confidence that it belongs to $f(x) \leq l_1$.
The objective is to use the output of the neural network to create a function, $\mathcal{N}_1(x)$, that returns the corresponding index of region.
See Eq.~\eqref{eq:actprobreg} for an example of such function.
As was discussed in Sec.~\ref{sec:dividedomain},
one can decide to make divisions with an objective in mind,
such as creating regions with similar size or contribution to variance.
When doing stratified sampling,
the usual criterion involves
dividing regions with larger variance.
This increases the freedom to optimize
the number of points per region, represented by $N_j$ in Eq.~\eqref{eq:SSVariance}.

One particularity of training networks for classification
is that data sets used for training
need to have similar sized classes.
Therefore,
if one of the regions becomes \emph{too small}
during the process outlined above,
it may be necessary to stop before reaching $\mathcal{C}_1$.
In this paragraph,
\emph{too small} is a rough way to say that the size of the sample in one region
may not be enough to achieve accurate predictions after training.
For a more precise description
one would need to consider the dimensionality of the function
and the complexity of the region that is being reduced.

\subsubsection{Example: finding divisions from the bottom}

The process above describes finding a single limit.
However, this process can continue in any of the newly created regions
and more regions can be obtained,
increasing the freedom to optimize the total variance.
In this section we will describe how one can
keep creating divisions by starting from the bottom
and using the classifiers from each step to keep moving to the top.
In this case $f_0$ would correspond to the minimum value of $f$.
If it is not known it can be guessed from the minimum value of $f(x)$
using the initial set $x\in\Phi$.
Precisely knowing the minimum is not critical because training is mostly concerned with limits that have points on both sides.
In fact, the minimum is only needed to calculate the first $\mathcal{C}(f_\text{min}, f_\text{test})$.
Assume that we have followed the process above and
have a function $\mathcal{N}_1(x)$ that outputs a predicted region index based on limit $l_1$.
We can use the following steps to go up the values of $f$ creating more divisions:
\begin{enumerate}
    \item
        Generate a large random set $L$ of points $x\in\Phi$, and calculate their predicted indices $\mathcal{N}_1(x)$.
    \item
        Select the points predicted in the upper region, i.e., those with higher confidence that they are above $l_1$.
        Let us call this set $\hat{L}_{1}$.
    \item
        Calculate $f(x)$ for $x\in \hat{L}_{1}$.
        It is not necessary to use the full $\hat{L}_{1}$,
        we just need enough to explore for the next division.
    \item
        From the previous step,
        if the trained model achieved good precision,
        we should have several points with $f(x) > l_1$.
        Use the values of $f(x) > l_1$ to follow the process of Sec.~\ref{sec:survey_single} for $\mathcal{C}(l_1, f_\text{test})$ with increasing $f_\text{test}$,
        up to step 3.
        This results in a new limit $l_2$.
    \item
        At this moment we should have $l_1$ and $l_2$.
        Add the set $\hat{L}_{1}$ (or the partial set used in step 4 for which we calculated $f(x)$)
        to the set used in the previous division.
        For this larger set create labels for all the points $x$ according to
        where $f(x)$ falls with the limits $l_1$ and $l_2$,
        i.e., three labels for three regions.
    \item
        Use the new labeled data to train a neural network
\end{enumerate}
The main featuren in this outline
is that we have used a priorly trained neural network
in selecting a set $x\in \hat{L}_{1}$,
where $\hat{L}_{1}$ would be the region
that the neural network learned is above $l_1$.
If the set $\hat{L}_{1}$ is considerably smaller than the full set $L$,
running $f$ only on $x\in \hat{L}_{1}$ should also be considerably faster
than running it on the full set $L$.
However, being realistic,
true improvement is achieved only if
running $f(x)$ for all $x\in L$
would take longer than the process of
training for $\hat{Y}_1$,
then getting $\hat{Y}_1(x)$ for all $x\in L$ and
then $f(x)$ for all $x\in \hat{L}_{1}$.
But also consider that a single training to get $\hat{Y}_1$ estimations
can be used for any size of $L$
as many times as necessary.
Therefore, the importance of time spent training is reduced with repeated usage of the same network.

These steps can be repeated, resulting in more limits on $f(x)$
and adding further output values to the next neural network,
having always one output value per intermediate level.

\subsection{Training the network to divide the domain}

The methodology is based on the premise that precise neural networks can effectively select points based on their function value $f(x)$ without the need to compute $f(x)$ directly. 
To enhance the efficiency of these neural networks, we employ an iterative training process as outlined in Ref.~\cite{Hammad:2022wpq}. 
This process seamlessly integrates with
Sec.~\ref{sec:survey}
by repeatedly predicting with new data sets, calculating true results and improving the training of the neural network.
Our approach focuses on accurately defining division boundaries, represented as $l_j$, within the neural network framework. 
This feature enables the classification of input points into their respective divisions based on their proximity to these boundaries. 
The important point of this methodology is to establish a consistent criterion for division assignment, thereby facilitating the accurate estimation of division lengths ($V_{\Phi_j}$) and the average function value within each division ($\langle f \rangle_{\phi_j}$).

The neural network is trained initially using a subset of points with known function values. 
This initial training phase aims to equip the neural network with the ability to classify points into their appropriate divisions. 
Once trained, the neural network predicts the division allocations for a larger, unclassified sample of points. 
These predictions are validated and corrected through direct computation of $f(x)$ for inaccurately classified points. 
The iterative nature of this process ensures continual refinement and retraining of the neural network, thereby enhancing its accuracy in division classification.

After achieving satisfactory accuracy, the neural network serves a dual purpose. 
Its first purpose is to allocate points to their respective regions.
This classification is essential for estimating the `length' or volume ($V_{\Phi_j}$) of each region, which is critical for understanding the spatial distribution of function values across the domain.
Second, the divisions established by the neural network allow for the estimation of average function values ($\langle f \rangle_{\phi_j}$) within each region, providing a comprehensive overview of the behavior of $f(x)$ across different divisions.
This estimation process prioritizes regions where the discrepancy between predicted and actual function values is significant, ensuring focused refinement in areas that most influence the overall accuracy of the neural network's performance.
In conclusion, this iterative training and application process of neural networks offers a robust framework for the efficient classification of points into defined regions based on the values of a given function. This approach not only enhances the precision of neural network predictions but also significantly contributes to the practical understanding and analysis of complex functions across varied domains. 
The purpose of the NN described here is displayed graphically in Fig.~\ref{fig:NNprocess}.

\subsection{Output activation function and performance}

\subsubsection{Output activation function and loss function}

The elements of the network setup
that define the output of the network
are the number of nodes of the output layer
and the output activation function.
As discussed previously,
the number of nodes depends greatly on the number of regions
and the activation function.
Since we have already discussed
the interplay between setting up the network to predict a number of regions,
here we want to discuss the another important element:
the loss function.
It is well known that after deciding on an activation function
there are typical choices for a loss function.
When using \emph{softmax} one tends to use \emph{categorical cross-entropy} (CCE),
in the case of \emph{sigmoid} one uses \emph{binary cross-entropy} (BCE)
while for \emph{tanh} the choice is the \emph{(squared) hinge} (SH) loss.
Considering that our intention
is to convert the network prediction
into guessed region indices,
the most natural improvement would be to include such information in the loss function.
We will take a simple approach for this,
where we take the difference between predicted and actual region index
and use it as weights in the loss function.
This a simple modification that can actually improve the accuracy of the model
and reduce the size of jumps between where a point $x$ is predicted and
where it belongs.
Let us use $f_\text{ltyp}$ to represent the typical loss function,
and the difference between predicted and actual region index
as $r_{\text{diff},j}$ with $j = 1,\ldots,N$, for $N$ training points.
We try two modifications of the loss function of the form
\begin{align}
    \label{eq:loss1}
    f_\text{loss1} & = \frac{1}{N}\sum_{j = 1}^N f_\text{ltyp}(\mathbf{Y}_j, \hat{\mathbf{Y}}_j) r_{\text{diff},j}^\alpha\\
    \label{eq:loss2}
    f_\text{loss2} & = \frac{1}{N}\sum_{j = 1}^N f_\text{ltyp}(\mathbf{Y}_j, \hat{\mathbf{Y}}_j)(1 + r_{\text{diff},j}^\alpha)
\end{align}
where the power $\alpha$ can be adjusted to improve convergence.
The arguments of $f_\text{ltyp}$, $\mathbf{Y}_j$ and $\hat{\mathbf{Y}}_j$, represent the target output and network output, respectively.
For the cases considered here,
the target and output of the network would be vectors.
Assuming vectorial output with $K$ elements, we consider the following forms of $f_\text{ltyp}$
\begin{align}
    \label{eq:cceloss}
    f_\text{CCE}(\mathbf{Y}_j, \hat{\mathbf{Y}}_j) & = -\sum_{k=1}^K Y_{j,k} \log (\hat{Y}_{j,k})\,,\\
    \label{eq:bceloss}
    f_\text{BCE}(\mathbf{Y}_j, \hat{\mathbf{Y}}_j) & = -\sum_{k=1}^K \left[Y_{j,k} \log (\hat{Y}_{j,k}) + (1 - Y_{j,k}) \log (1 - \hat{Y}_{j,k})\right]\,, \\
    \label{eq:shloss}
    f_\text{SH}(\mathbf{Y}_j, \hat{\mathbf{Y}}_j) & = \sum_{k=1}^K (1 -Y_{j,k} \hat{Y}_{j,k})^2\,.
\end{align}
For CCE,
the expression corresponds to the usual case
where output is first one-hot encoded,
also known as multiclass.
In the case of BCE and SH we consider a multilabel scenario
where each element of $\mathbf{Y}_j$
represents whether the result of the calculation
is above or below each limit,
as was mentioned in Sec.~\ref{sec:nnoutputindices}.
In the case of BCE, we would use \emph{sigmoid} as output activation function for each element of $\mathbf{Y}_j$,
while for SH we would have \emph{tanh}.
For the two last cases,
we train using \emph{sigmoid} (\emph{tanh}) with labels 0 ($-1$) and 1 (1)
for below and above each limit, one \emph{sigmoid} (\emph{tanh}) for each considered limit.
In Sec.~\ref{sec:nnoutputindices}
we have mentioned how to convert this type of output to region indices.

As we will see in the next section,
these modifications to loss functions sometimes bring instability,
related to $r_{\text{diff},j}$ moving by integer values.
In some cases,
the loss function can become more stable
by multiplying Eqs.~\eqref{eq:loss1} and~\eqref{eq:loss2}
with the typical loss function
in the form $f_{\text{loss}1,2(\times)} =f_{\text{loss}1,2}\times \sum_{j=1}^N f_\text{ltyp}(\mathbf{Y}_j, \hat{\mathbf{Y}}_j)/N$.
This creates a balance between including and not including weighting by integer $r_{\text{diff},j}$.

\subsubsection{Performance}

To test performance with the proposed modified loss functions
and corresponding output activation function,
we train 10 times with each combination and compare the resulting tendencies.
As metrics to determine which combinations perform better,
we use accuracy defined as fraction of points correctly classified,
and average jumping defined as the sum of all the differences
between predicted region and correct region, divided over the number of tried points.
We choose these two metrics based on the possibility of a highly accurate network
where the misclassified points have large jumps ultimately increasing variance
and complicating sampling with the neural network.

\begin{figure}
    \centering
    \includegraphics[width=0.45\textwidth]{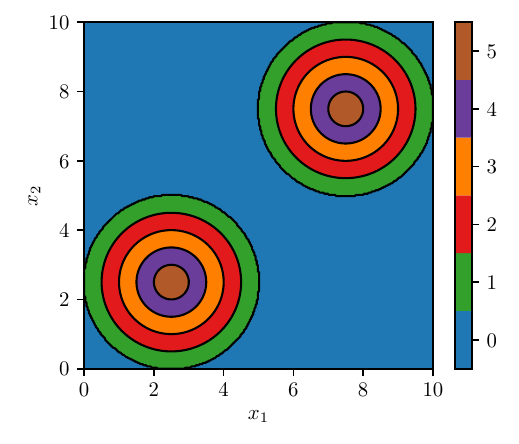}%
    \includegraphics[width=0.45\textwidth]{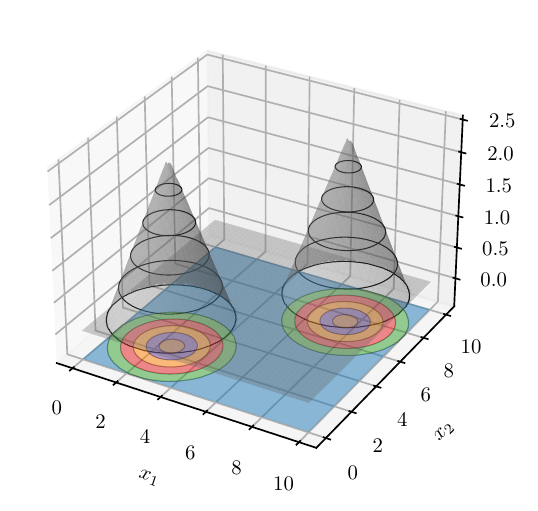}%
    \caption{\label{fig:cones2d}%
        Example of a two-cones function with a 2-dimensional base
        divided into 6 regions labeled from 0 to 5.
        The bases of the cones are centered at (2.5, 2.5) and (7.5, 7.5)
        with a maximum height of 2.5.
        Regions labeled 1 to 5 correspond to regions limited
        by $\{0, 0.5, 1.0, 1.5, 2.0, 2.5\}$
        while the region labeled 0 is for $f_\text{cones} = 0$.
        The panel on the left shows color coded regions
        and limits projected on the space of the two-dimensional base.
        The panel on the right shows the cones in 3-dimensional space,
        with darker lines to visualize how the limits are applied on $f_\text{cones}$.
        Note that this example is for visualization.
        See the text for a detailed description of a test with a 6-dimensional base
        divided in 12 regions.
    }
\end{figure}

As test function,
we use a combination of cones
with $d$-dimensional base.
The height of each cone is given by a function $f_j(\vec{x})$
with $\vec{x}$ a $d$-dimensional vector and
\begin{equation}
    f_{\text{cone-}j}(\vec{x}) = %
    \begin{dcases}
        r_j - |\vec{x} - \vec{c}_j| & \text{if}\quad  |\vec{x} - \vec{c}_j| < r_j \,, \\
        0 & \text{if}\quad  |\vec{x} - \vec{c}_j| \geq r_j \,,
    \end{dcases}
\end{equation}
where $r_j$ and $\vec{c}_j$
are the maximum height and center of the cone,
respectively.
For a single cone,
the regions would be concentrical rings around $\vec{c}_j$
at different heights of $f_j(\vec{x})$,
except for the region where $f_j = 0$
which would be a $d$-cube minus a $d$-sphere of radius $r_j$.
In order to add complexity to the task of learning regions with a neural network,
we consider a combination of two separated cones with $\vec{x}$ given in 6 dimensions
\begin{equation}
    f_\text{cones}(\vec{x}) = f_{\text{cone-}1}(\vec{x}) + f_{\text{cone-}2}(\vec{x})
\end{equation}
with
\begin{align}
    r_1 & = r_2 = 2.5 \,, \\
    \vec{c}_1 & = 2.5 \times \vec{1} \,,\\
    \vec{c}_2 & = 7.5 \times \vec{1} \,,
\end{align}
where $\vec{1}$ is a vector with all 6 entries equal to 1
and the space is limited to $[0, 10]$ in all dimensions.
In this case all regions are given by two disconnected rings,
except the region where $f_\text{cones} = 0$
which is a 6-cube
minus two 6-spheres of radius 2.5.
The regions are labeled 0 to 11,
with regions 1 to 11 limited by
the contours
$f_\text{cones} = \{$0.0,
0.012,
0.094,
0.181,
0.278,
0.387,
0.513,
0.663,
0.852,
1.125,
1.801,
2.5$\}$.
These limits where chosen such that $V^2_{\Phi_j} \sigma^2_{\Phi_j} (f_\text{cones})$ in all regions
are roughly the same.
The 0$^\text{th}$ region is chosen as $f_\text{cones} = 0$.
To aid visualization,
a very simple example with a 2-dimensional base
and 6 regions is shown in Fig.~\ref{fig:cones2d}.

To test performance of loss functions,
we employ a simple multilayer perceptron
with two hidden layers.
For all the architectures considered in this section,
the input layer receives the six dimensional vector $\vec{x}$
and the two hidden layers have 1920 and 960 nodes with ReLU activation function.
The size of the output layer
depends on the number of regions
we want to classify.
As mentioned earlier,
we divide $f_\text{cones}$ in 12 regions.
This means that for multiclass,
using \emph{softmax} activation, the output is 12 dimensional.
For multilabel, where we use either \emph{sigmoid} and \emph{tanh} activations,
the output is an 11 dimensional vector.
An easy way to see why the multilabel approach requires one less output value
is to remember that for two labels
classification can be done with a single output.
Then add one more output for each extra label.

\begin{figure}
    \centering
    \includegraphics[width=0.45\textwidth]{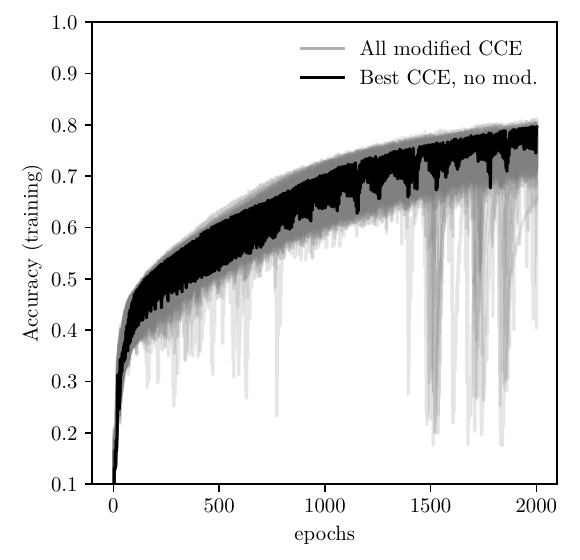}%
    \includegraphics[width=0.45\textwidth]{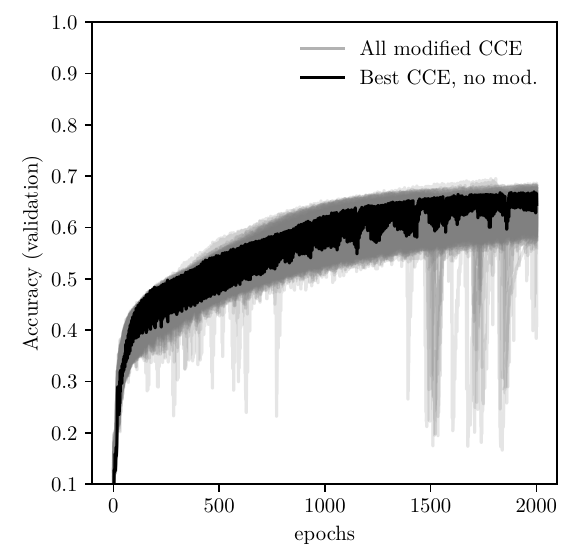}
    \includegraphics[width=0.45\textwidth]{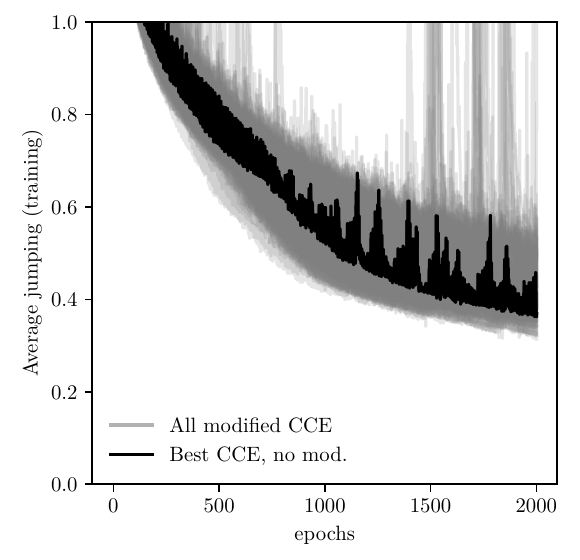}%
    \includegraphics[width=0.45\textwidth]{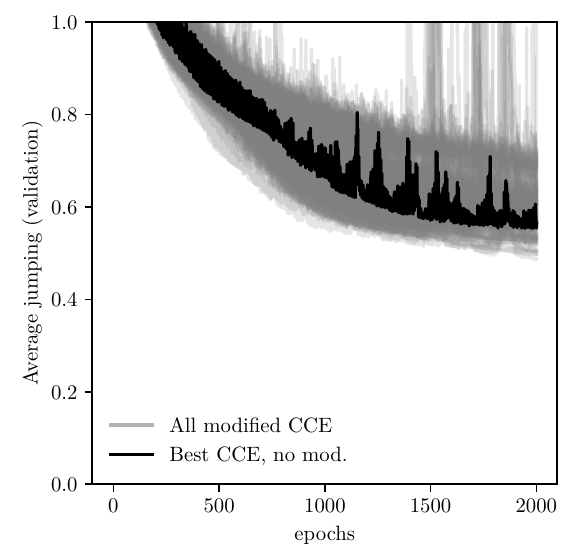}
    \caption{\label{fig:ccemod}%
        Evolution of two metrics, accuracy (top row) and average jumping between regions (bottom row),
        for the combination of \emph{softmax} output activation function
        and categorical cross-entropy (CCE) loss.
        In the text this is described as multiclass approach.
        The training is done for 2000 epochs.
        The panels on the left show metrics on the training set
        while panels on the right show metrics on the validation set.
        For CCE without modification only the best out of 10 trainings is shown (black),
        while 10 trainings with several modifications (described in the text) are shown in gray.
        None of the proposed modifications brings a considerable improvement of the metrics.
    }
\end{figure}

\begin{figure}
    \centering
    \includegraphics[width=0.45\textwidth]{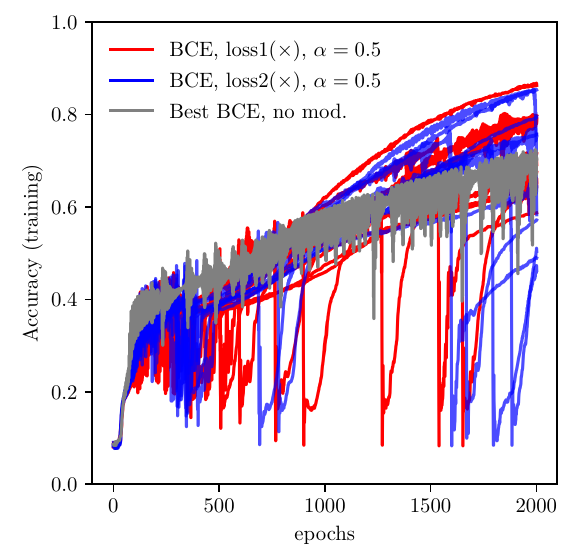}%
    \includegraphics[width=0.45\textwidth]{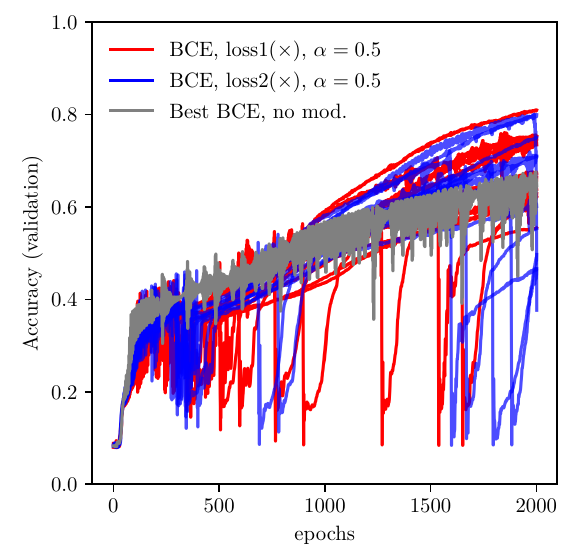}
    \includegraphics[width=0.45\textwidth]{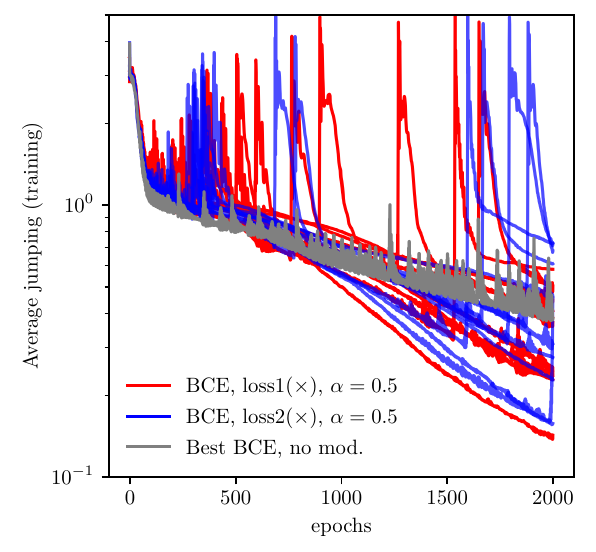}%
    \includegraphics[width=0.45\textwidth]{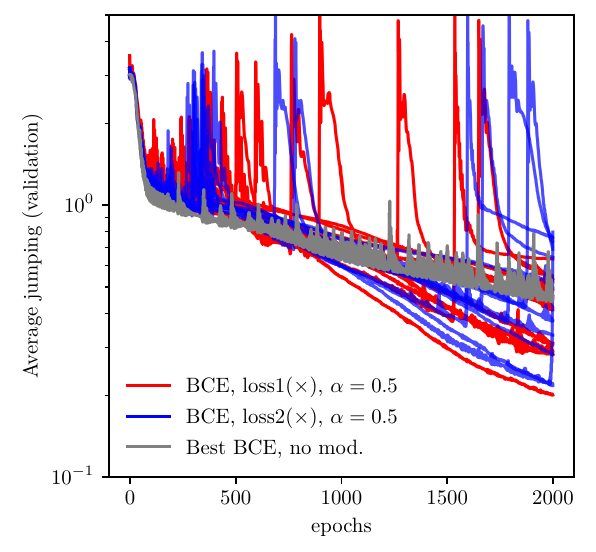}
    \caption{\label{fig:bcemod}%
        Evolution of two metrics, accuracy (top row) and average jumping between regions (bottom row),
        for the combination of \emph{sigmoid} output activation function
        and binary cross-entropy (BCE) loss.
        In the text this is described as multilabel approach.
        The training is done for 2000 epochs.
        The panels on the left show metrics on the training set
        while panels on the right show metrics on the validation set.
        For BCE without modification only the best out of 10 trainings is shown (gray),
        while 10 trainings for the two best performing modifications are shown in red and blue.
        None of the proposed modifications brings a considerable improvement of the metrics.
    }
\end{figure}

\begin{figure}
    \centering
    \includegraphics[width=0.45\textwidth]{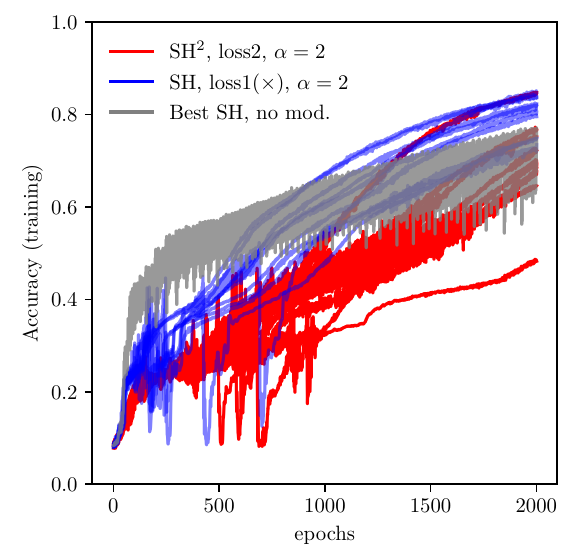}%
    \includegraphics[width=0.45\textwidth]{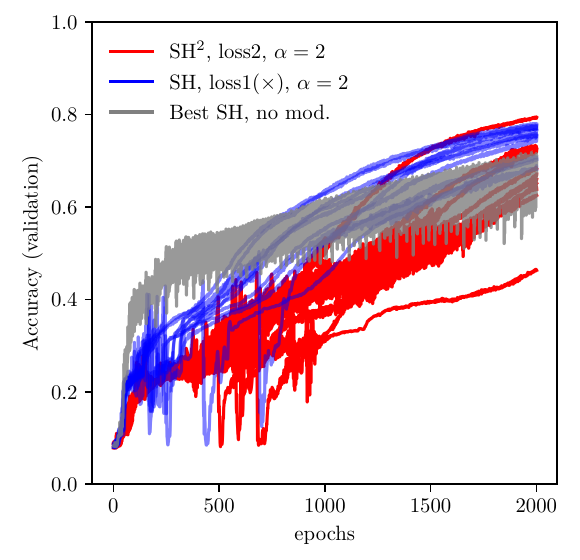}
    \includegraphics[width=0.45\textwidth]{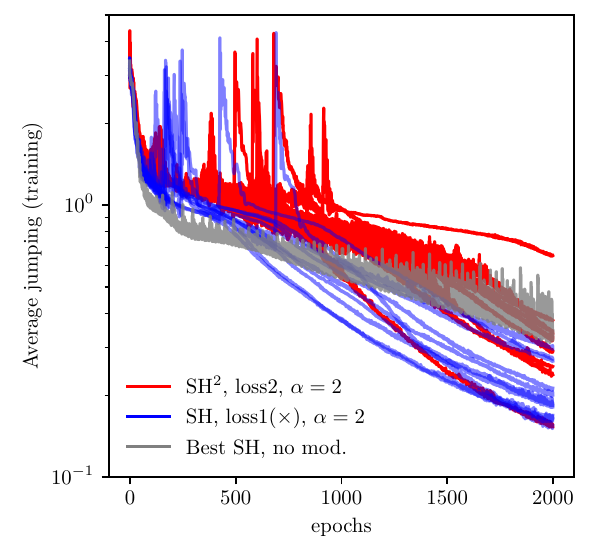}%
    \includegraphics[width=0.45\textwidth]{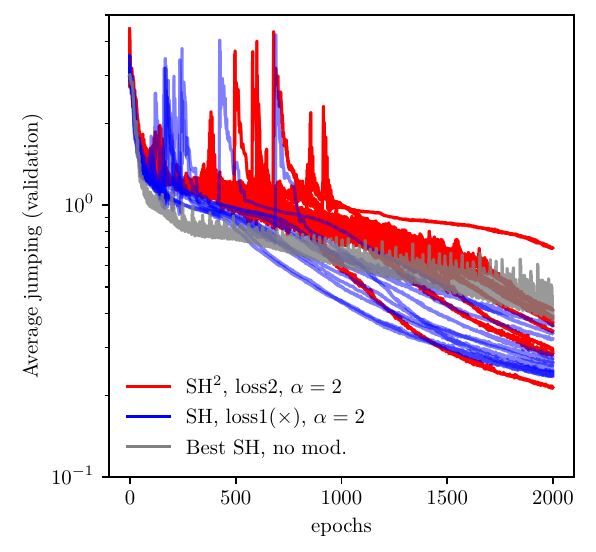}
    \caption{\label{fig:shmod}%
        Evolution of two metrics, accuracy (top row) and average jumping between regions (bottom row),
        for the combination of \emph{tanh} output activation function
        and squared hinge (SH) loss.
        In the text this is described as multilabel approach.
        The training is done for 2000 epochs.
        The panels on the left show metrics on the training set
        while panels on the right show metrics on the validation set.
        For SH without modification only the best out of 10 trainings is shown (gray),
        while 10 trainings for the two best performing modifications are shown in red and blue.
        None of the proposed modifications brings a considerable improvement of the metrics.
    }
\end{figure}

To test performance of the different combinations of network configuration and loss function,
we apply 10 different trainings for 2000 epochs and compare the evolution of accuracy
and average jumping between regions (misclassification distance).
For training we employ 6000 points and set aside another 6000 points for validation.
The results of our tests are displayed in Figs.~\ref{fig:ccemod} to~\ref{fig:shmod}
for typical loss functions and modifications with comparable to better performance.

In Fig.~\ref{fig:ccemod},
we demonstrate that the proposed modifications to CCE
do not bring any significant improvement to any of the metrics.
The modifications include both types of loss described in Eqs.~\eqref{eq:loss1} and~\eqref{eq:loss2}
with $f_\text{ltyp} = f_\text{CCE}$ and
three different $\alpha = \{0.5, 1.0, 2.0\}$.
It can be seen that in some cases the result becomes unstable.
In the case of BCE and modifications,
Fig.~\ref{fig:bcemod} shows that modifications can perform better
increasing accuracy and reducing average jumping between regions.
Ten trainings are shown for two modifications
that show better performance compared
to a single training, the best out of 10, with BCE without modifications.
The two modifications correspond to Eqs.~\eqref{eq:loss1} (red) and~\eqref{eq:loss2} (blue)
multiplied by the typical BCE, with $\alpha = 0.5$.
While they improve the metrics shown in Fig.~\ref{fig:ccemod},
they turn out to be very unstable.
Such instability can be somewhat alleviated by conditionally stopping training
via callbacks.
For SH and modifications,
results are shown in Fig.~\ref{fig:shmod}
for two modified loss functions compared against
the best out of 10 trainings using SH without modifications.
The blue lines show 10 trainings using Eq.~\eqref{eq:loss1} with $alpha = 2$
and multiplied by SH to improve stability.
The red lines show 10 trainings using Eq.~\eqref{eq:loss2}
but this time the typical loss function has the form of Eq.~\eqref{eq:shloss}
with a power of 4 instead of 2.
Both options show improvement in the evolution of the metrics after 1000 epochs,
with the blue lines showing the most consistent results.
Although some instability is displayed before 1000 epochs,
all runs seem to become stable after that.
When comparing all the options
an interesting feature is that the disagreement between training and validation sets
is more pronounced in CCE (and modifications) than when using BCE and SH (and their modifications).
Moreover,
CCE and its modifications bring the least reduction in average jumping compared to the other
options considered.

\begin{figure}
    \centering
    \includegraphics[width=0.33\textwidth]{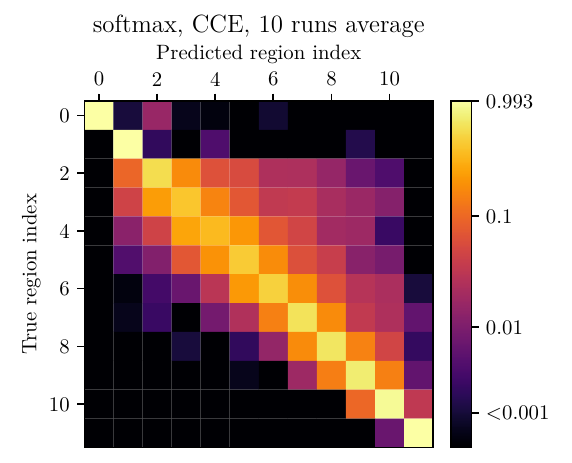}%
    \includegraphics[width=0.33\textwidth]{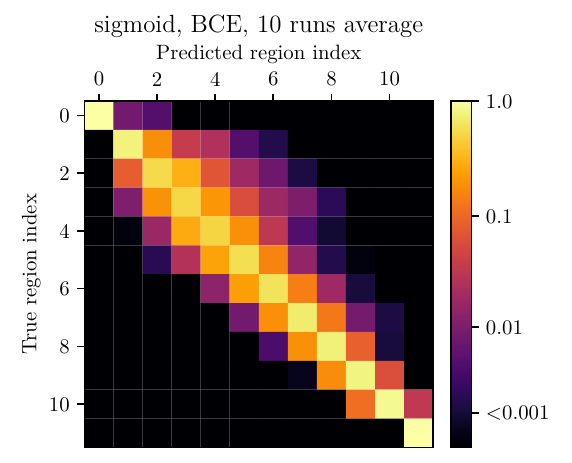}%
    \includegraphics[width=0.33\textwidth]{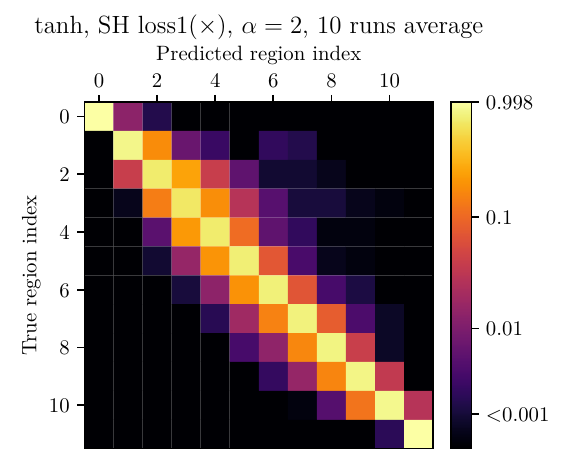}
    \caption{\label{fig:confmatcompare}%
        Average confusion matrices for 10 trainings with approximately 3000 epochs.
        The values of each matrix have been normalized to total number of points in true region index.
        Results are shown only for loss functions that yielded the best 10-runs average
        for three different activation functions of the output layer.
        The combination of activation function and loss function
        are indicated on the top of each figure.
    }
\end{figure}

Finally,
we employ confusion matrices to paint a more detailed representation of the degree of misclassification.
In Fig.~\ref{fig:confmatcompare} we show confusion matrices for the three options of activation function
and the best performing loss function.
The results shown correspond to the average of 10 different training with approximately 3000 epochs.
An early stopping callback was added
to stop training when average jumping between regions did not improve considerably,
thus ameliorating some of the effects of unstable loss functions.
On the upper left panel we can see that the combination of \emph{softmax} and CCE
performed very well for the extreme regions but poorly for middle regions.
The combination of \emph{sigmoid} and BCE is shown on the upper right panel.
The default BCE performed better than modified versions that presented high variation in 10 runs due to
unstable trainings.
We can see that, on average, the network is more accurate for higher regions, after 8,
while below 8 it tends to misclassify upwards.
In the case of \emph{tanh}, the modified version corresponding to $f_\text{loss1}$ with $\alpha = 2$
and multiplied again by the default SH loss gave the best performance of all.
This can be interpreted from the almost totally diagonal confusion matrix in the lower panel of Fig.~\ref{fig:confmatcompare}.
For this reason, in the examples that will be tested in the rest of this work we will
mostly employ \emph{tanh} as activation function in the output layer and some variation of SH.
One common characteristic among the three confusion matrices
is that the most accurate classification
is achieved for the extremal regions.
This may be related to misclassifications in extremal regions happening only in one direction,
while for middle regions misclassifications can go in both directions.
Note that there is no guarantee that the performance exhibited in this section
will be consistent among different problems.
This section mostly demonstrates that
judiciously choosing a combination of activation and loss functions
may lead to improved performance.

To conclude this part,
we acknowledge that the function employed in this test is by no means complicated
when compared to results from calculations performed in a proper study.
However, in this section we are more interested in comparing performance
between networks with different outputs trained using different loss functions.
The characteristics of the 2-cones function that are more appealing for this test
are the presence of several regions, most of them with disconnected sections
and the presence of regions that are completely surrounded by other regions.

\subsection{Rationale behind this machine learning algorithm}

To summarize the role of ML in our approach,
we aim to integrate two crucial pieces of information into the neural network: identifying the region to which a point belongs and understanding the hierarchical relationships between these regions. 
Particularly, embedding hierarchy into a neural network is key to ensuring that misclassified points fall into nearby regions, thereby reducing prediction variance and enhancing the model's robustness.

To explain the importance of the neural network in variance reduction,
let us bring here the expression for variance of integral estimate in a single region,
from Sec.~\ref{sec:fullvariance}
\begin{align}
    \sigma^2(E(I_j))
    \label{eq:fullvarapprox2}
    \approx E^2(V_{\Phi_j}) \sigma^2(\langle f \rangle_{\Phi_j})
    + E^2(\langle f \rangle_{\Phi_j}) \sigma^2(V_{\Phi_j})
    + \sigma^2(V_{\Phi_j})\sigma^2(\langle f \rangle_{\Phi_j})\,.
\end{align}
The point of stratified sampling
is to make non-overlapping regions of the domain of the integral
and optimize the number of points required in each
subdomain according to their variance.
To understand how variance is reduced
in the process explained above and for each region,
let us describe what happens term by term:
\begin{itemize}
    \item
        $E^2(V_{\Phi_j}) \sigma^2(\langle f \rangle_{\Phi_j})$:
        the first term is reduced, independently of the process,
        by increasing the number of regions.
        In our case,
        recall that $\sigma^2(\langle f \rangle_{\Phi_j}) = \sigma^2_{\Phi_j}(f)/N_j$,
        then increasing the regions
        directly reduces $\sigma^2_{\Phi_j}(f)$
        by constraining $f(x)$ to a smaller range.
    \item
        $E^2(\langle f \rangle_{\Phi_j}) \sigma^2(V_{\Phi_j})$:
        the second term is specific of cases where volume of regions has to be estimated.
        This term can have large contributions due to $E^2(\langle f \rangle_{\Phi_j})$,
        that can only be balanced by reducing the variance of the estimated volume, $\sigma^2(V_{\Phi_j})$.
        As we pointed out in Sec.~\ref{sec:MCfsize},
        region volumes, $V_{\Phi_j}$, can be estimated
        with the ratio of points found inside a region
        over the total number of sampled points.
        This estimation can be done relying solely on the neural network
        and, therefore,
        $\sigma^2(V_{\Phi_j})$ is reduced by evaluations of the neural network.
        With a fast random numbers generator and a performant neural network
        it is possible estimate $V_{\Phi_j}$ to great accuracy
        independently of how expensive or time consuming $f(x)$ may be.
    \item
        $\sigma^2(V_{\Phi_j})\sigma^2(\langle f \rangle_{\Phi_j})$:
        This term automatically goes down when the other two are reduced.
        In fact, if we require that $\sigma^2(\cdot)/E^2(\cdot) < 1$,
        this term becomes negligible much faster than the other two.
\end{itemize}

It is clear that the second term can become specially problematic
for integrands with large values,
most notably in cases where the target integral value is comparably smaller.
Without the neural network to reduce variance in the second term,
we would need to run $f(x)$ to classify points into regions
and then estimate volumes.
For regions with large $E^2(\langle f \rangle_{\Phi_j})$
this could potentially require far more $f(x)$ evaluations than those
required to reduce the first term.
This can be taken as a reasonable justification
for any procedure that divides the integral domain
in regions of known $V_{\Phi_j}$ where $\sigma^2(V_{\Phi_j})$ vanishes.
For our process, it is the employment of a neural network
what brings potential advantage for functions
that may be slow to evaluate.
Summarizing, 
the first contribution to variance
is reduced by choosing smaller ranges,
while the second term is reduced by evaluations of the neural network.
Thus, we have effectively moved the most difficult task
to the neural network that is faster to evaluate.

\section{Integration with LeStrat-Net}
\label{sec:examples}

\subsection{One dimensional function with three peaks}

\begin{figure}[htb]
    \centering
    \includegraphics[scale=0.6]{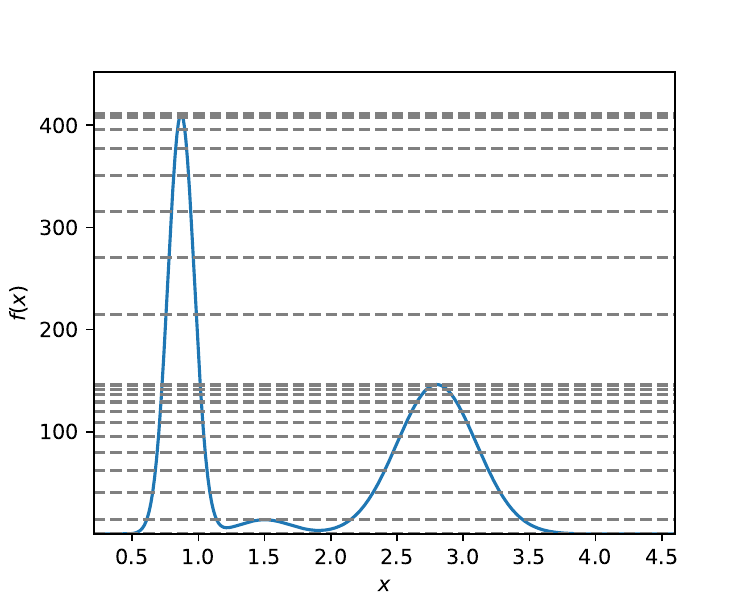}
    \includegraphics[scale=0.6]{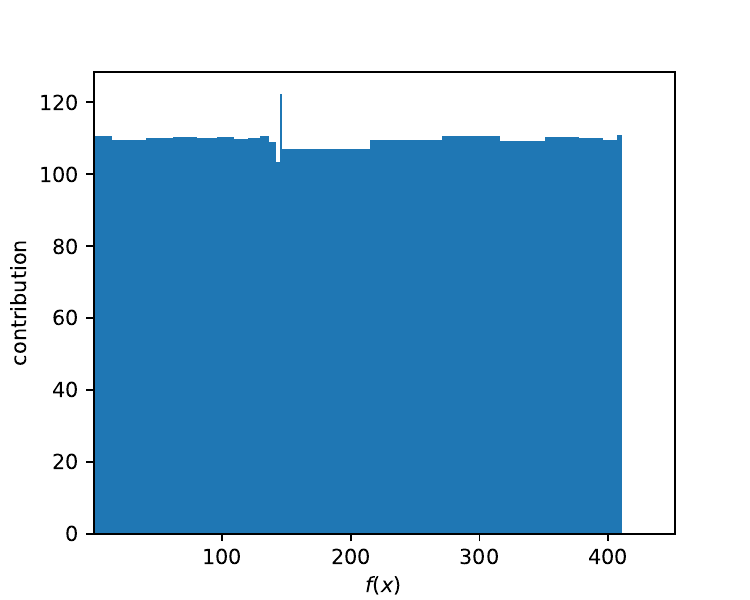}
    \includegraphics[scale=0.6]{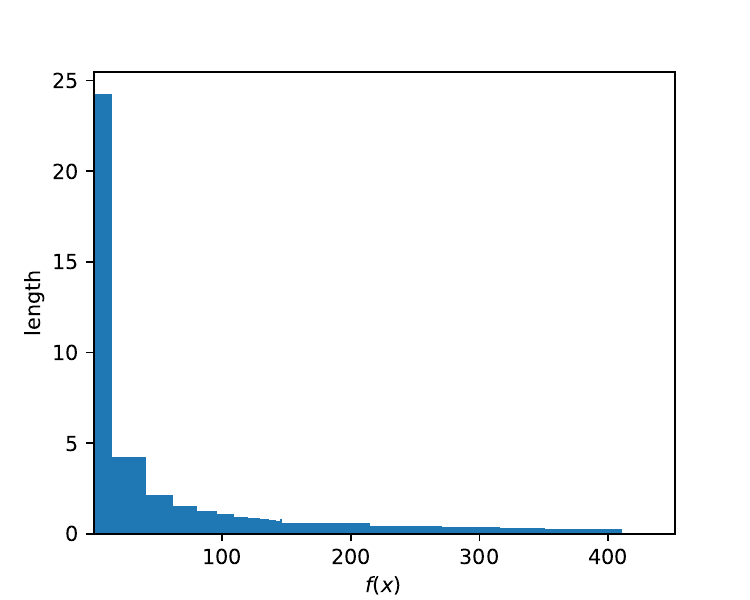}
    \includegraphics[scale=0.6]{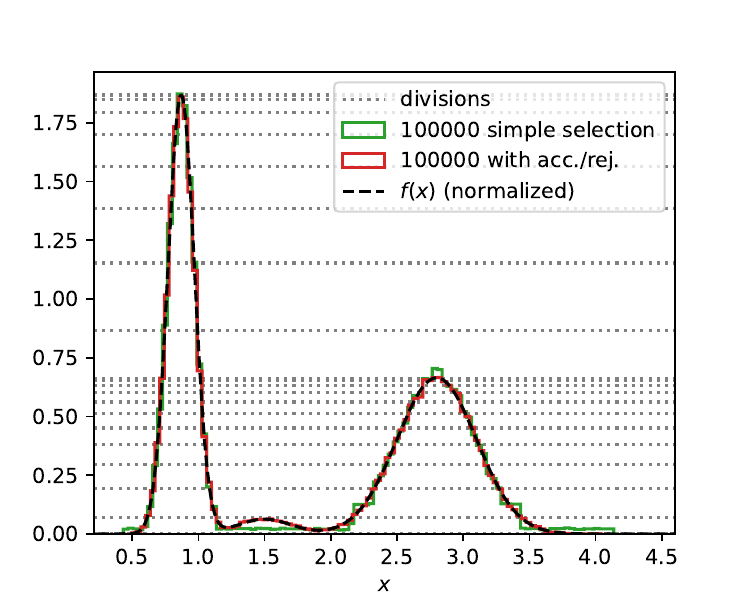}
    \caption{\label{fig:threepeaks}%
        Simple example with a function with three peaks.
        The upper left panel shows the function in solid blue
        and the limits of the divisions with dashed gray lines.
        The upper right panel shows that the 20 divisions have roughly the same contribution.
        In the lower left panel is shown the resulting $x$-length
        from projecting the sections in the $x$-axis.
        The lower right panel shows an attempt to generate 100\,000 points
        following the distribution of $f(x)$.
        The divisions on $f(x)$ are shown as dotted lines.
        100\,000 samples generated by simply selecting points in each region
        according to contributions are represented as a normalized histogram in green,
        while the 100\,000 samples of the red histogram use acceptance/rejection.
        The function $f(x)$, normalized to 1, is shown as a black dashed line.
    }
\end{figure}

To illustrate the idea presented above,
consider a 1-dimensional function with three peaks with different heights.
The values of the function are shown in the upper left panel of Fig.~\ref{fig:threepeaks}.
It corresponds to the sum of three gaussian functions:
centered at 0.87, 1.5  and 2.8; with deviations 0.1, 0.2 and 0.3, respectively.
The three gaussians are initially normalized and later multiplied by
103, 7, and 110.
A total of 20 divisions are made from $f(x) = 10^{-5}$
to $f(x) \approx 411$, with each division
contributing roughly the same to the area under the curve.
The lengths that results from projecting the divisions
in the $x$-axis are shown in the lower panel,
showing that that regions with higher value $f(x)$
have smaller lengths even though they contribute comparatively similar to other
larger regions.

Using this information we can select a number of points, $K$,
distributed according to the contributions to the area under the curve
shown in the upper-right panel Fig.~\ref{fig:threepeaks}.
We can do this selection by simply choosing an appropriate number of points, $K_j$,
from each section from a sample of points.
The number of points is related to the contribution of each section, $I_{\Phi_j}[f(x)]$,
to the full integral
\begin{equation}
    K_j = K \frac{I_{\Phi_j}[f(x)]}{I_\Phi [f(x)]}
\end{equation}
This was done for the green histogram of the lower right panel of Fig.~\ref{fig:threepeaks}.
In the case of regions where $f(x)$ changes over a wide range of $x$,
some corners may appear below or above the target distribution,
while for regions where $f(x)$ changes more sharply
the adjustment has better accuracy.
In the case of the highest peak, the tip is missed by the green histogram.
This could be corrected by further dividing larger sections
or by generating the sample already distributed proportionally as $f(x)$.
Another procedure involves applying acceptance/rejection of points,
such that points with larger $f(x)$ have a larger acceptance 
than points with lower $f(x)$.
In our case we can take advantage of the fact that we set divisions
that limit the value of $f(x)$ in each section.
For example, take the $j^\mathrm{th}$ division, limited by $f(x) = l_{j + 1}$,
give to each point $x_i \in \Phi_j$ a confidence $c_{fj}(x) = f(x_i)/l_{j + 1}$.
Even if $x_i$ is uniformly distributed,
since we can rescale by $l_{j + 1}$, we can reach large acceptance
regardless of small $l_{j + 1}$ values.
The red histogram in the lower right panel of Fig.~\ref{fig:threepeaks} shows an example of using
acceptance-rejection to select 100\,000 points.

\subsection{Oscillating function, region contours and network size}
\label{sec:oscillating}

\begin{figure}[tb]
    \centering
    \includegraphics[width=0.5\textwidth]{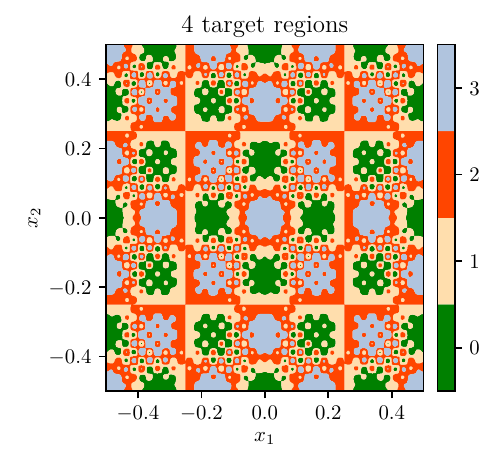}
    \caption{\label{fig:foscregs}%
        4 regions for the two dimensional oscillatory function $f_\text{osc}$
        defined as described in the text.
        Each different color indicates a single region.
    }
\end{figure}

One example of a difficult function
that also helps illustrate
two interesting details of this method of stratification,
is a highly oscillating function.
In this section,
we want to consider a function
that presents a combination of sine and cosine functions with different frequencies.
For easy visualization
we consider a two dimensional function of the form
\begin{equation}
    \label{eq:fosc}
    f_\text{osc}(x_1, x_2) = \prod_{j=1}^2\sin(40\pi x_j) \sin(4\pi x_j) + \prod_{j=1}^2\cos(6\pi x_j)\,,
\end{equation}
in the range $x_j = [-0.5, 0.5]$.
This function presents an oscillatory behavior
that will result in regions with several disconnected subregions
of different sizes.
For Eq.~\eqref{eq:fosc} 4 regions create shapes that are complicated enough,
for this test, as can be seen in Fig.~\ref{fig:foscregs}.
the regions are defined by
\begin{align}
    \label{eq:oscreg0}
    \Phi_0:&\quad f_\text{osc} \leq -0.5\,, \\
    \label{eq:oscreg1}
    \Phi_1:&\quad -0.5 < f_\text{osc} \leq 0\,, \\
    \label{eq:oscreg2}
    \Phi_2:&\quad 0 < f_\text{osc} \leq 0.5\,, \\
    \label{eq:oscreg3}
    \Phi_3:&\quad f_\text{osc} > 0.5\,.
\end{align}

Rather than checking that we can repeat every single feature of the regions,
we will concentrate
on reducing the difference in variance
between true and predicted regions.
This reduction in variance
is reflected in the number of points
required for integration.
The reasoning behind this is that
missing important features would result in larger variance.
Conversely, trying to fit too many details
increases the requirements during training
such as higher number of network nodes, more epochs
and more points required for training.
Obviously, when presented with regions with complicated shape,
a larger network will result in a more accurate prediction.
Similarly, training with more points
results in a more accurate network,
however,
here we want to train with the least possible number of points.
For this test we want to check the effect of network size
and number of training points in accuracy of predictions.
We will test fully connected neural networks with two hidden layers.
For all networks the input layer is two dimensional
and the output is three dimensional with \emph{tanh} activation function.
We will consider a network that we will label ``small''
with 80 and 40 nodes in each hidden layer,
with ReLU activation function.
We will compare this with another network,
labeled ``large'',
with 800 and 400 nodes in each hidden layer.
These two networks will be trained following the iterative
process described in Ref.~\cite{Hammad:2022wpq},
mostly paying attention to correcting wrong predictions.
We will also consider a variant of the ``large'' network,
referred to as ``large+restart'',
that will be trained from zero
with different sizes of training sets,
but following the same prescription for selection of points
as for the other networks.
The training will use 3000 epochs for each network,
but for the ``large+restart'' we will also test a longer training of 30\,000 epochs
with the same final set as the ``large'' network.
For all the examples the loss function will be \emph{squared hinge}.

\begin{figure}[htb]
    \centering
    \includegraphics[width=0.45\textwidth]{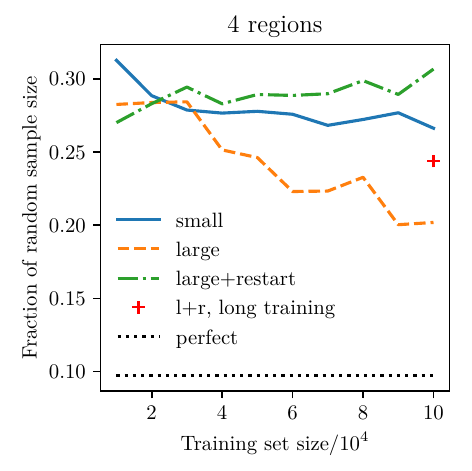}
    \caption{\label{fig:foscreduc}%
        4 regions: Reduction in sample size required to achieve a certain error of integration.
        This reduction is defined as the sum of number of samples required
        from each of the 4 regions to achieve a certain error,
        over the number of points required from a random uniform sampling
        (no divisions) to achieve the same error.
        The number of samples used for training is shown in the $x$-axis.
        The ``large+restart'' network with long training is only trained with
        the maximum size of training set used and is shown as a red cross.
        For all cases 3\,000 epochs are used, except for ``l+r, long training''
        where we used 30\,000.
    }
\end{figure}

To compare performance
we will consider the fraction of sample size
required to achieve a certain error
when using stratification over having no divisions.
For this purpose,
let us assume that the number of points per region
is proportional to $V_{\Phi_j}^2 \sigma_{\Phi_j}^2(f)$.
In that case,
for a given variance of estimates $\sigma^2_{E}$,
the required number of points per region is
\begin{equation}
    \label{eq:njreq}
    N_j = \frac{n_\text{reg} V_{\Phi_j}^2 \sigma_{\Phi_j}^2(f)}{\sigma^2_E}\,,
\end{equation}
and the total number of required points from stratified sampling is
\begin{equation}
    \label{eq:nstratreq}
    N_\text{strat} = \frac{n_\text{reg}}{\sigma^2_E} \sum_j V_{\Phi_j}^2 \sigma_{\Phi_j}^2(f)\,.
\end{equation}
From the expression above for $n_\text{reg} = 1$ and a single region
we obtain the required points when not applying stratification, $N_\text{single}$.
Divide $N_\text{strat}$ over $N_\text{single}$ and we obtain the fraction of points required
by applying stratification
\begin{equation}
    \frac{N_\text{strat}}{N_\text{single}} = \frac{n_\text{reg}}{V_{\Phi}^2 \sigma_{\Phi}^2(f)}\sum_j V_{\Phi_j}^2 \sigma_{\Phi_j}^2(f)\,.
\end{equation}
By dividing in 4 regions,
as in Eqs.~\eqref{eq:oscreg0} to~\eqref{eq:oscreg3},
the fraction is around 0.1.
In other words, if we had a way to perfectly select samples
from each of the defined 4 regions,
we would need to evaluate $f_\text{osc}$ ten times less
than if we drew samples from a random uniform distribution of the whole space.
The networks can provide an approximation to this selection of samples,
with additional evaluations of $f_\text{osc}$ to obtain training samples.
The results for this reduction for different networks with different sizes of training sets
is shown in Fig.~\ref{fig:foscreduc}.
Expectedly,
the ``large'' network (orange dashed line) has the best evolution
of the three cases considered,
starting at a reduction of around 0.28
down to $\sim 0.2$ of the sample size required
when no divisions are used.
It is also the only case
that keeps consistently giving a lower fraction of samples
when increasing the size of the training set.
The ``small'' network (blue solid line) also shrinks the required sample
but does not show a significant improvement
after a training set of 30\,000 points.
The case of a large network trained every time from zero,
labeled ``large+restart'' (green dot-dashed line) actually has the worst evolution,
with the size of required samples actually increasing
when training with a larger training set.
The variant of this where we train the large network
from zero for 30\,000 epochs and with the 100\,000 training set (red cross)
improves performance of network prediction,
but does not result in a reduction
as low as the ``large'' case.
This indicates that performing progressive trainings
with increasing training set,
as described in \cite{Hammad:2022wpq},
presents an improvement when compared with long trainings
with large training sets.
For comparison, we show the perfect case as a dotted black line.

\begin{figure}[htb!]
    \centering
    \includegraphics[width=0.45\textwidth]{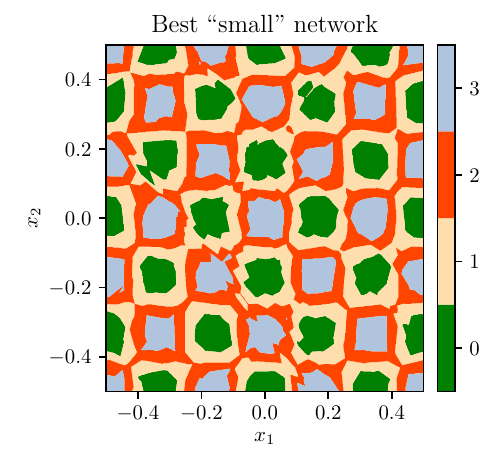}%
    \includegraphics[width=0.45\textwidth]{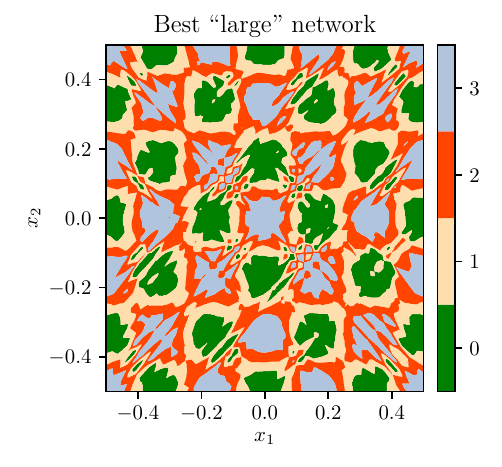}
    \includegraphics[width=0.45\textwidth]{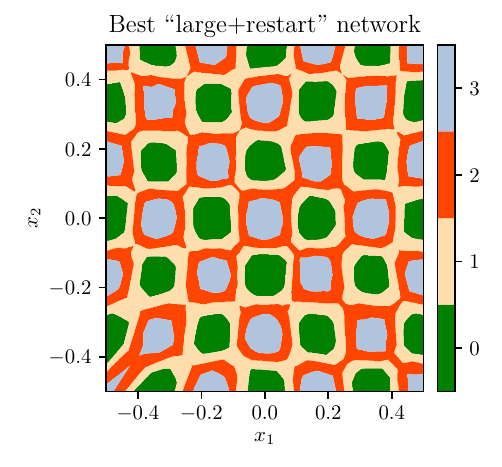}%
    \includegraphics[width=0.45\textwidth]{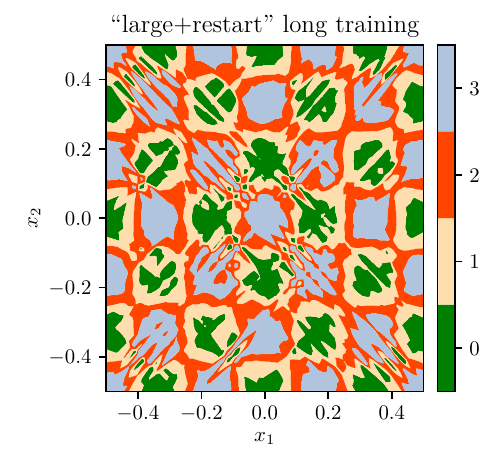}
    \caption{\label{fig:foscnetmaps}%
        Maps of regions obtained from the best performing networks
        for the four cases considered in Sec.~\ref{sec:oscillating}.
        This corresponds to the minimum fraction obtained
        in each line of Fig.~\ref{fig:foscreduc},
        except for the bottom-right panel for which there is only one point.
        The upper-left panel corresponds to the ``small'' network case, with 100\,000 training points.
        The upper-right panel is for the ``large'' network with 90\,000 training points,
        the lower-left panel shows the map for the large network trained from zero,
        with 10\,000 training points.
        The lower-right panel shows the map for a network of the same size as the ``large'' case
        with a training set of 100\,000 points trained for 30\,000 epochs.
    }
\end{figure}

In Fig.~\ref{fig:foscnetmaps}, we show maps of the predicted regions
using the networks from each case with the lowest fraction in Fig.~\ref{fig:foscreduc}.
For the ``small'' case (upper-left) this corresponds to the network
trained with 100\,000 points.
However, considering the improvement in the ``small'' case
is not obvious after 30\,000 training points
one can expect a similar prediction after this value.
In the upper-left panel of Fig.~\ref{fig:foscreduc}
it is seen that smaller features of regions
tend to be assigned to contiguous regions.
For the larger features,
all minimum and maximum regions have been properly predicted by the network.
For the ``large'' network case (upper-right) we show the network
trained with 90\,000 points.
In this case we can see that the network is able to predict
more intricate regions,
managing to predict some fine features closer to the center.
When going far from the center,
we see that the network recognized some of the details that are aligned
with the diagonals.
For the variant where we try a network with the ``large'' architecture
but trained from zero every time,
we display the training with 10\,000 points in the lower-left panel.
Posterior trainings performed worse.
However, similarly to the ``small'' network case,
largest features such as minimums and maximums have been all correctly placed.
Considering that this corresponds to the first training,
we can expect that the first iteration of the ``large'' case
also started with a similar prediction and evolved to the figure in the upper-left panel.
Finally, the result obtained from a network similar to the ``large'' case
but trained from zero with 100\,000 points during 30\,000 epochs is shown
in the lower-right panel.
Visually, this network also attempts to repeat finer features of regions,
with larger details shown around the center of the figure.
While it may not be obvious from Fig.~\ref{fig:foscnetmaps},
the upper-right figure performs better than the lower-right.
This is demonstrated in Fig.~\ref{fig:foscreduc},
where the reduction obtained with the upper-right map is $\sim 0.2$
while for the lower-right is $\sim 0.25$.

\begin{figure}[htb]
    \centering
    \includegraphics[width=0.45\textwidth]{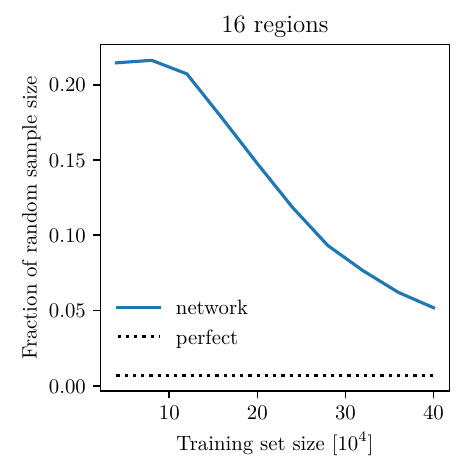}
    \caption{\label{fig:fosreduc16}%
        16 regions: reduction of the fraction of required points for integration
        by training the network to learn 16 regions
        via iterative training with different sizes of training set.
        The perfect case ($\sim 0.007$) is shown as a black dotted line.
    }
\end{figure}

While integration can certainly be performed with 4 regions,
a larger number of divisions is required for a more competitive result.
For this purpose, we increase the number of regions to 16,
where, in the perfect case, the fraction of points required goes down to 0.007.
This corresponds to 16 regions with similar sized $V_{\Phi_j} \sigma_{\Phi_j}(f_\text{osc})$,
with limits
$\{-1.065$, $-0.847$, $-0.665$, $-0.497$, $-0.359$, $-0.231$, $-0.108$, 0.002,
0.106, 0.230, 0.359, 0.498, 0.664, 0.838, $1.067\}$
used analogously to Eqs.~\eqref{eq:oscreg0} to~\eqref{eq:oscreg3}.
We also increase the size of the network accordingly.
Following the example of the large network,
we follow the same template but change the number of nodes in the two hidden networks
to 3200 and 1600, and the number of output nodes to 15,
one output per limit as before.
We also increase the training epochs to 4000,
and train initially with 40\,000 points and add 40\,000 at every refined training step,
up to 400\,000 training points.
The reduction of the fraction of required points for integration
after every training
is shown on Fig.~\ref{fig:fosreduc16}.
While the perfect case is not achieved,
the best performing network achieves a fraction of $\sim 0.052$.

\begin{figure}[htb]
    \centering
    \includegraphics[width=0.45\textwidth]{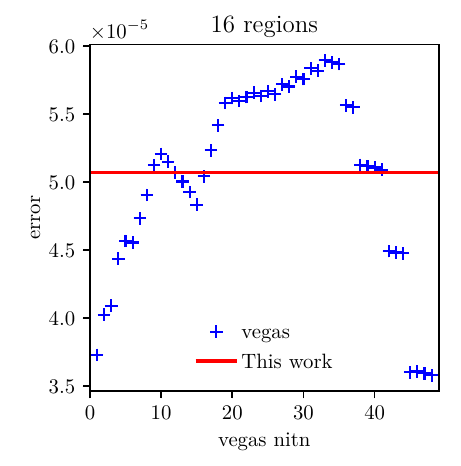}%
    \includegraphics[width=0.45\textwidth]{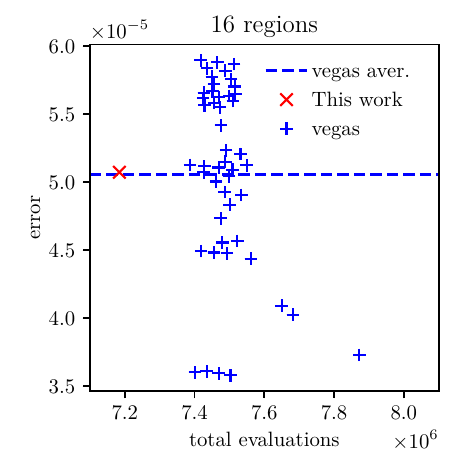}
    \caption{\label{fig:fosinteg16}%
        16 regions: Results of performing integration using the network to classify regions.
        The results from \texttt{vegas} are shown in blue
        and the error from 100 integrations with our method
        is shown as a red line of the left panel and a red cross on the right panel.
        The left panel shows the error obtained from 48 \texttt{vegas} runs
        with different number of iterations (\texttt{nitn}).
        The red horizontal line
        indicates the error from 100 integrations with the network for $\sim 7.18\times 10^6$ evaluations (training and integration).
        The right panel shows the same points
        according to the total number of evaluations of $f_\text{osc}$.
        The dashed blue line indicates the average error of the 48 \texttt{vegas} runs
        (see text for more information.)
    }
\end{figure}

With the aim of evaluating the competitiveness of our integration,
we integrate $f_\text{osc}$ with \texttt{vegas}~\cite{Lepage:2020tgj},
for different values of the number of iterations, \texttt{nitn},
and reaching a total number of evaluations around $7.7\times 10^6$.
In the case of \texttt{vegas} we find that the error remains around $~5\times 10^{-5}$.
Limiting \texttt{nitn} to the range 1-48,
we make sure to include points with larger and lower error,
as can be seen in the left panel
of Fig.~\ref{fig:fosinteg16},
with \texttt{nitn} above 48 giving larger error.
For our result with 16 regions,
we perform 100 integrations with an estimated error of $5.05\times 10^{-5}$,
consistent with the average error of the \texttt{vegas} runs.
We find that in our case we need $~7.2\times 10^6$ evaluations
of $f_\text{osc}$ to hit the average error from \texttt{vegas},
as shown in the right panel of Fig.~\ref{fig:fosinteg16}.
In the same figure it can be seen that the number of evaluations
used for \texttt{vegas} integrations are above but not far away from this value.
Note that this example is not a demonstration of superiority
but of comparable performance.
In this case the resulting errors and number of evaluations are not considerably different
and, more importantly, Monte Carlo methods tend to display advantageous performance
in higher dimensionality.
A more detailed comparison with more dimensions is presented in the next section.

To summarize the results from this section,
predicting intricate regions may require larger networks,
something that should already be expected.
However, smaller networks may be able to predict broader features
of regions, hiding the details into contiguous regions.
The iterative way of training the network resulted in better performance,
even when compared to the same network trained once with the largest training set
for an equivalent number of epochs.
For the 4 regions case, even after the first training,
all networks started with a reduction of sample size
of $\sim 0.3$,
which may not be close to the perfect case $\sim 0.1$
but still represents and advantage against random uniform sample.
In the case of performing integration,
considering more regions further reduces the sample size
but requires a larger network.

\subsection{Function with large cancellation}

\begin{figure}[tb]
    \centering
    \includegraphics[width=0.5\textwidth]{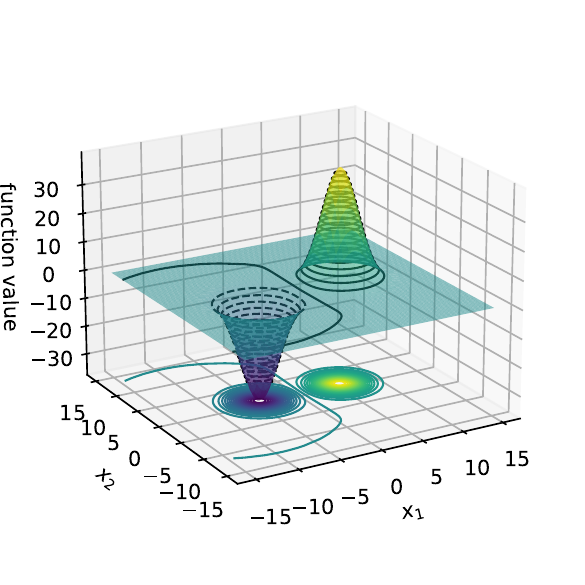}
    \caption{\label{fig:fcancel}%
        Example of a function with large cancellation
        with dependence on two coordinates.
    }
\end{figure}

The point of any MC based method is to make calculations in high dimensionality.
With this in mind,
here we consider an example of a function with large cancellation
in 7 dimensions.
An example of a function with cancellation is shown in Fig.~\ref{fig:fcancel}.
Additionally, rather than only showing that the process above works in 7 dimensions,
we will take the popular MC method \texttt{vegas} as a reference
in order to compare against a well established method.

The function will be constructed with multidimensional normalized gaussians
with some factors to control their contribution to integral value.
Let us call $g(x; \mu, \sigma)$ to the normalized gaussian function for a single coordinate $x$.
This function is given by
\begin{equation}
    \label{eq:gaussiansingle}
    g(x; \mu, \sigma) = \frac{1}{\sigma \sqrt{2\pi}} \exp\left[
        \frac{(x - \mu)^2}{2\sigma^2}
    \right]\,.
\end{equation}
To construct seven dimensional gaussians we take the product of functions of this type
applied individually to each of the 7 components of the vector $\vec{x}$
\begin{equation}
    G(\vec{x}; \vec{\mu}, \sigma) = \prod_{j=1}^7 g(x_j; \mu_j, \sigma)\,,
\end{equation}
and combine three of them into
\begin{equation}
    \label{eq:fcancel7d}
    f_{7D} (\vec{x}) = 100\times \left[
        G(\vec{x};\vec{\mu}_+, \sigma_+) - G(\vec{x};\vec{\mu}_-, \sigma_-)
     \right] + 0.1\times G(\vec{x}; \vec{0}, \sigma_0)
\end{equation}
where $\vec{0}$ is a 7-dimensional vector full of 0s.
The values of $\vec{\mu}_+$ and $\vec{\mu}_-$
are decided randomly
with the condition that $\vec{\mu}_+$ must be all-positive values
while $\vec{\mu}_-$ is all-negative values.
The values used for the test described here correspond to:
\begin{align}
    \vec{\mu}_+ & = (1.429, 0.838, 1.580,
     0.304, 1.729, 1.270, 1.543)\,, \\
    \vec{\mu}_- & = (-1.997, -1.464, -1.775,
     -0.765, -1.096, -0.622,
     -1.694
    )\,,\\
    \sigma_+ & = \sigma_- = 0.3\, \\
    \sigma_0 & = 1.0\,.
\end{align}
From Eq.~\eqref{eq:fcancel7d},
it is clear that integrating $f_{7D}$
over the space $(-\infty,\infty)^7$ results in 0.1.
Here we limit ourselves to the range [$-5$,5] for each of the 7 dimensions,
which is wide enough to effectively take 0.1 as the expected result from integration.

The determination of limits follows the process of Sec.~\ref{sec:survey},
targeting regions of similarly sized
$V_{\Phi_j} \sigma_{\Phi_j}(f_{7D})$.
This means that, with the same number of evaluations of $f_{7D}$,
all regions would have a similar contribution to total variance
from the first term of Eq.~\eqref{eq:fullvarapprox}.
Besides making new divisions based on this criterion,
if some regions end up with $V_{\Phi_j} \sigma_{\Phi_j}(f_{7D})$ 4 orders of magnitude smaller than average,
we will merge such regions
with the contiguous regions with the smallest
$V_{\Phi_{j\pm 1}} \sigma_{\Phi_{j\pm 1}}(f_{7D})$.
The reason for this is to reduce the number of regions
that the networks needs to learn.
Another point to take care while making divisions,
is that a limit may be set
even if the criterion above has not been met
if, otherwise, the region may not have enough data
to properly training the network.
In summary: limits should create regions
of similar $V_{\Phi_j} \sigma_{\Phi_j}(f_{7D})$
without creating others for which not enough data has been accumulated.
As more data is accumulated,
more accurate $V_{\Phi_j} \sigma_{\Phi_j}(f_{7D})$ can be calculated
and is easier to create regions with enough training data.

\begin{figure}[tb]
    \centering
    \includegraphics[scale=0.7]{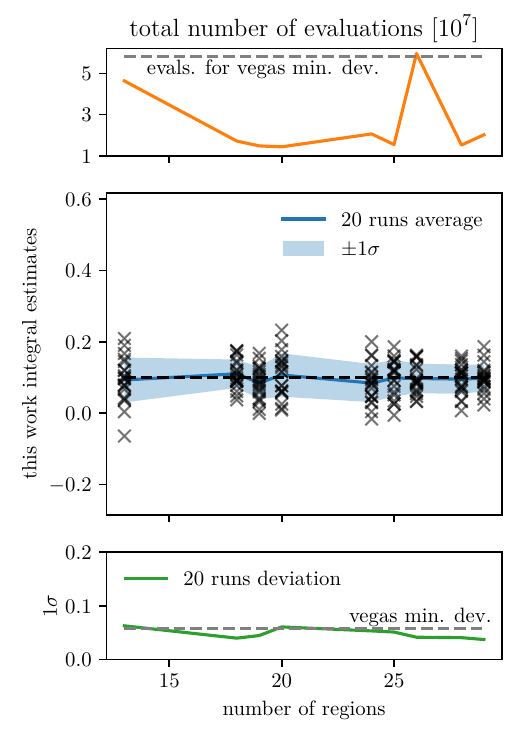}%
    \includegraphics[scale=0.7]{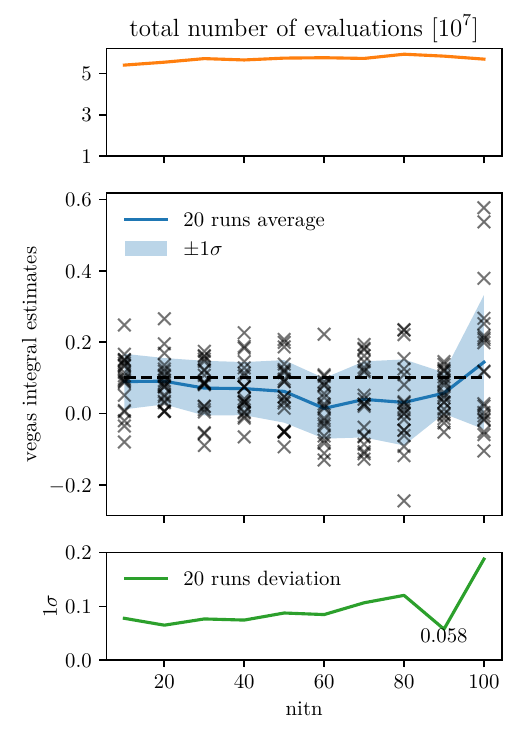}
    \caption{\label{fig:vegas_vs_ml}%
        Results for several integration attempts for a function with big cancellation in 7 dimensions.
        The top row shows the number of evaluations of $f_{c7D}(x)$,
        the center shows results for the estimation of the integral
        and the bottom row shows the $1\sigma$ deviation for 20 runs.
        For our approach (left) the $x$-axis corresponds to number of regions.
        A case of a particularly bad training is shown for 26 regions.
        In the case of \texttt{vegas} (right) we vary the number of iterations (\texttt{nitn})
        while trying to keep the number of evaluations of the same size.
        On the left, the dashed gray line represents the result from the right
        with the least deviation (0.058).
        The dashed black line represents the target value of the integral (0.1).
    }
\end{figure}

Regarding the network setup,
we use a fully connected neural network
with two hidden layers.
The input to the network is $\vec{x}$ but normalized to the range [$-1$,1].
For predictions,
we use the multilabel approach
described in Sec.~\ref{sec:nnoutputindices}.
The output activation function is \emph{tanh}
with as many output nodes as number of regions minus 1.
The number of nodes in the first hidden layer
is twice the number of regions times the number of dimensions,
while for the second hidden layer is half of that.
This configuration is not particularly special
and it is possible that more or less nodes or layers
may perform better.
For optimization of the network,
we use the Adam algorithm
with a learning rate of 0.001
and \emph{squared hinge} as loss function.
For training we use 5000 points per region.
Note that a new network has to be trained
when new limits are added,
However, new limits may be added for several regions
before training a new network.
For this test,
we decide to stop the creation of regions
at different steps,
to compare performance with different number of regions.
To evaluate performance of integration,
we use the final network
to collect a number of points per region
enough to reduce contributions from the first term of Eq.~\eqref{eq:fullvarapprox}
down to a total of 0.5.
Contributions from the second term of Eq.~\eqref{eq:fullvarapprox}
are reduced by passing random points to the network
and counting the number of points in each region,
until such contribution is 0.1 of the contribution from the first term.
The third term automatically becomes negligible.
We perform this integration 20 times
and use those 20 estimates to calculate the error of estimation.
The results are shown in the left panel of Fig.~\ref{fig:vegas_vs_ml}.
In the top part we show the total number of evaluations of $f_{7D}$,
including the number of evaluations required during training and determination of limits of regions.
We see that for most of our tests, the number of required evaluations
is below $5\times 10^7$, becoming lower for larger number of regions.
There is a peak at 26 regions,
due to an inaccurate network
that requires more points for final integration.
It is expected that for some number of regions the number of required evaluations should grow due to the increase in required points for training.
However, such growth seems to be above 29 regions.
In the middle we see the estimates of the integral,
with the average shown as a blue line and the $\pm 1\sigma$ error
from the 20 estimates shown in light blue.
We can see that the average follows the line of the correct result,
shown as a dashed black line,
even for the inaccurate network.
This means that even if something goes wrong with training,
we still have access to the actual result,
at the cost of more evaluations of $f_{7D}$.
In the bottom, we show in green the value of the $1\sigma$ error
from the 20 runs.
This error remains around 0.5-0.6 which is consistent
with our request that integration stops when the total error from the first term of Eq.~\eqref{eq:fullvarapprox} is at 0.5.

To compare against \texttt{vegas},
we ran 20 different integrations for values of
the number of iterations, \texttt{nitn},
from 10 to 100, in steps of 10.
The number of evaluations per iteration is set such that
integration requires around $6\times 10^7$ evaluations of $f_{7D}$.
Instead of relying on error estimate from \texttt{vegas},
we use the 20 integrations to estimate the error of the integration estimates.
The results are shown in the right panel of Fig.~\ref{fig:vegas_vs_ml}.
The most notable feature
is how the average of estimates tends
to underestimate the integral for \texttt{nitn} between 20 to 90,
with the average closer to 0 around 60.
For \texttt{nitn} around 100 and larger
the average integral estimate recovers
but the error goes up,
as expected from having to distribute the $\sim 6\times 10^7$
among more iterations.
For $\texttt{nitn} = 90$, the 20 integrations present the least deviation
and we will use this as a reference in our result.
In the left panel of Fig.~\ref{fig:vegas_vs_ml},
we show as a gray dashed line,
the result from \texttt{vegas} with the least deviation.
In the case of the number of evaluations,
we can see that most of our results are well below the gray dashed line,
except for the inaccurate network that requires a similar amount of evaluations.
For the $1\sigma$ error, we see that all of our results are at a similar level
or below.
Note that our result is shown across number of regions
while \texttt{vegas} is shown across values of \texttt{nitn},
which cannot be compared directly.
The most important detail in this comparison,
is how we consistently estimate the value of integrand
across number of regions
while keeping a lower number of evaluations
for most of our attempts.

\begin{figure}[t]
    \centering
    \includegraphics[scale=0.6]{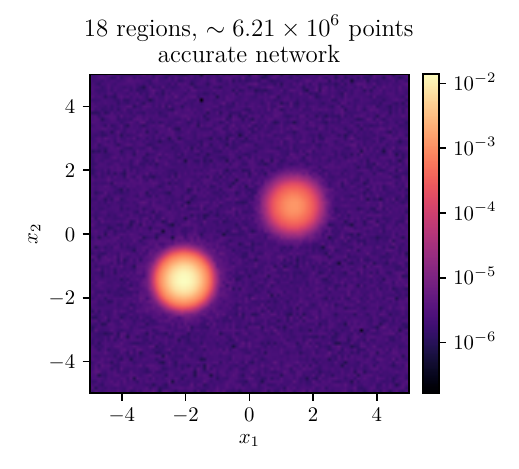}%
    \includegraphics[scale=0.6]{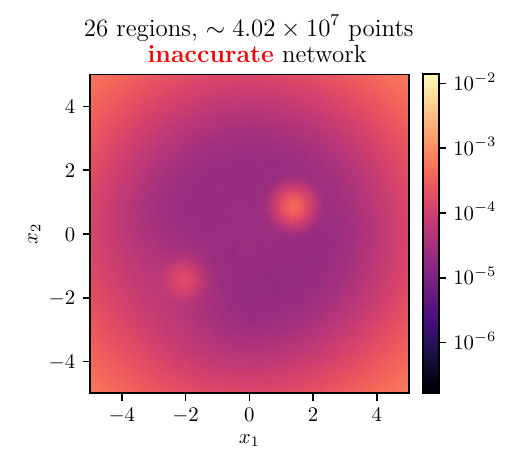}%
    \includegraphics[scale=0.6]{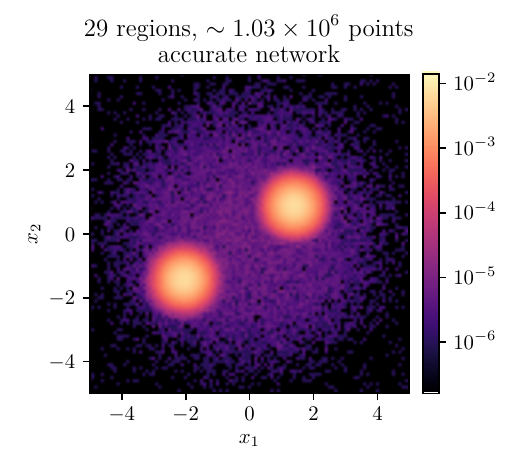}
    \caption{\label{fig:bcancelhist2d}%
        Two dimensional projection
        of the normalized density of the sample
        selected by the neural network.
        These densities correspond to points
        that were selected purely from network prediction
        and used during the last integration step.
        Larger number of regions
        can capture more details for accurate networks (negligible misclassification by more than 1)
        except in cases where the network is inaccurate (excess of misclassification by more than 1).
        The same color map is used for the threes figures with black representing density below the scale of the color map including 0.
    }
\end{figure}

To see the effect from having more or less regions,
in Fig.~\ref{fig:bcancelhist2d} we show 2D histograms
of the number of points used for final integration,
projected in the first two dimensions.
We can see that for 18 regions,
two peaks, corresponding to the positive and negative peaks,
are used to calculate the cancellation,
while a uniform background is used for the rest of the space.
For the inaccurate network, with 26 regions,
we can see that the two peaks are recognized
but are reduced in importance
by oversampling around the borders of the space.
In the case of 29 regions,
we can see the effect of having more regions and an
accurate network.
For this example, we can see the two peaks clearly,
but also in the background we can see the shape of the
smaller gaussian.

\section{Generation of scattering events}

In high energy particle physics,
the most common use of Monte Carlo simulations
is in the generation of events for collider physics studies.
The basic approach
is to generate a large number
of kinematically allowed points
for some $n_i \to n_f$ process
where $n_i$ and $n_f$ stand for number of initial and final particles.
Then, we can assign a weight to each generated point by evaluating
squared amplitudes and parton distribution functions (PDF).
In general, the weight of some phase space point $p$ is given by
\begin{equation}
    w(p) = \prod_{j}^{n_i} f_j(x_j, \mu) \left|\mathcal{M}_{n_i \to n_f}(p)\right|^2 \left|J(p)\right|
\end{equation}
where $f_j$ represents PDFs
for incoming particles with momentum fraction $x_j$ and
factorization scale $\mu$,
$\mathcal{M}_{n_i \to n_f}$ is the matrix element of the process,
and $J$ stands for any Jacobian factor in case a change of variables is applied.

Assume we want to generate $N_{ev}$ events,
the outline of the process is given by
\begin{enumerate}
    \item Determine the relative contribution of each region,
    which is proportional to $V_{\Phi_j} \langle f \rangle_{\Phi_j}$.
    \item Based on this relative contribution,
    determine the share that each region will contribute to $N_{ev}$.
    \item Generate phase space points and use the neural network to decide
    which region they belong to.
    \item In each region, obtain the weight of the obtained phase space points
    and apply acceptance/rejection to unweight the events.
    Stop when the number of required events has been completed.
\end{enumerate}
Increasing the accuracy of the network and with narrower divisions,
the efficiency of the acceptance rejection step increases for each region,
independently of the region having lower or larger height.

\subsection{Generating events for $q\bar{q} \to e^+ e^-$}

To demonstrate the event generation process described above,
we generate 10$^5$ collider events for the process $u\bar{u} \to e^+ e^-$.
In this 2-to-2 process,
the number of free parameters consists on the energy of the incoming $u$ quarks
and the 3-momentum of one of the outgoing leptons.
The rest of the kinematical variables can be obtained from those three by using
conservation of momentum and energy.
The collisions are assumed to happen on a 13~TeV collider,
therefore, the quarks have a maximum energy of 6.5~TeV,
with their distribution given by the corresponding PDF.
To calculate PDFs we use LHAPDF 6.5~\cite{Buckley:2014ana} with sets from NNPDF 2.3~\cite{Ball:2012cx}.
The amplitude squared is calculated using MadGraph 3.5~\cite{Alwall:2014hca}
using the standalone mode to generate fortran source code
that is later compiled into a \texttt{python} module using \texttt{f2py}.

As mentioned in other examples,
we use a fully connected neural network with two hidden layers.
The first hidden layer has $250\times$number of regions nodes,
while the second one has half of that.
The input layer receives 5 dimensional input for the five
independent values that can be used to reconstruct the event:
two up-quark energies and 3-momentum for either positron or electron.
The output layer has as many nodes as number of regions minus 1
using \emph{tanh} activation function.
The loss function used is \emph{squared hinge},
and we optimize using the Adam algorithm
with a learning rate of 0.001,
using 4000 epochs per training
with 3 retrainings.
We start with two regions but keep redividing regions with larger variance.
Every time we add more regions we also increase the number of retrainings
by 3.
In each retraining we add wrongly classified points
from the pool of points that have already been calculated.
For training we use 5000 points per region.

\begin{figure}[tb]
    \centering
    \includegraphics{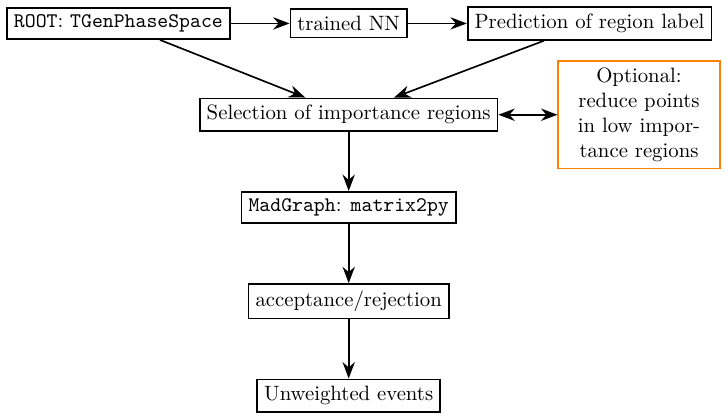}
    \caption{\label{fig:nnevgen}%
        Generation of events with selection of regions based on NN prediction.
    }
\end{figure}

The generation of events described in this section
is shown in the chart of Fig.~\ref{fig:nnevgen}.
The phase space is generated using \texttt{TGenPhaseSpace} from \texttt{ROOT}~\cite{Brun:1997pa}
and the network classifies each point generated into its predicted region.
Points are the selected according to importance
and run through the matrix element calculator to obtain their weight.
After the acceptance/rejection step, unweighted events are obtained.

\begin{figure}
    \centering
    \includegraphics[scale=0.7]{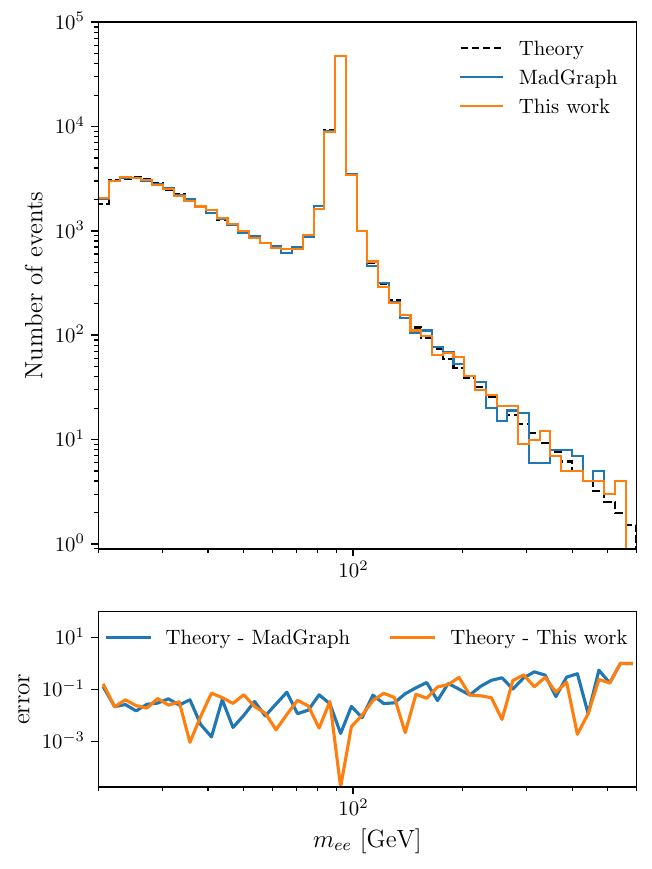}%
    \includegraphics[scale=0.7]{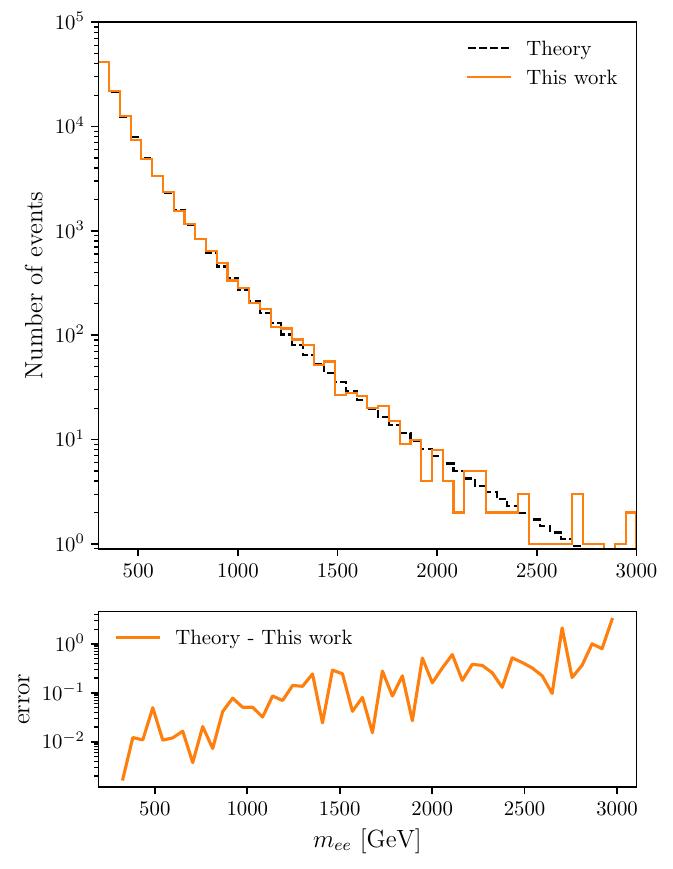}
    \caption{\label{fig:qqee_events}%
        Histogram for 10$^5$ events generated using the method described in this work (solid blue)
        and with MadGraph's default settings (orange).
        The error is the difference with the theoretical results which is displayed as a black dashed line.
    }
\end{figure}

The results from selecting 100\,000 for the $u \bar{u} \to e^+ e^-$ example
are shown in Fig.~\ref{fig:qqee_events}.
The histogram for our generated events is shown in orange
and is compared against the theoretical prediction (black dashed line)
and the histogram from the same number of MadGraph generated events.
Below the histogram we show the error obtained with our method
and with MadGraph, with both being somewhat consistent,
although our result has the lowest error around the peak of the $Z$ resonance,
as can be seen on the bottom left panel.
In the right panel, we show results from using the network to generate events 
in a higher $m_{ee}$ invariant mass region,
consistent with region of lower importance.

Now that we have demonstrated generation of events
with our approach, we comment on possible extensions.
Thinking on the example of generating events in low importance regions,
by using the approach described in this work
it is possible to concentrate on regions regardless of their 
height in importance.
Specially, regions that hold interesting features,
even if their importance is lower,
may be studied independently of other regions.

\section{Conclusions}
\label{sec:conclusion}

In this work we have proposed a new approach to divide the domain of the integrand function
when doing stratified Monte Carlo.
We divide the domain using isocontours of the result
of a calculation,
effectively performing Monte Carlo sampling
in an style analogous to Lebesgue integration.
In this way, regions are defined by their position in output space
and can have any shape or even be composed by disconnected regions.
We combine this with a ML approach
to have a neural network that learns the complicated shapes of the regions
and can preclassify large amounts of data without having to run the full calculation on each point.
The neural network becomes the ultimate divider of the domain space
and can be used to obtain more information on the regions
such as estimation of volumes and the variance of this estimation.

We present all the necessary ingredients to this purpose,
including comparisons of performance, modifications to loss functions,
instructions on how to convert the output of the neural network into classification of multiple regions
and how to formulate different criteria for the placement of divisions.
The creation of divisions that limit the range of the output of the calculation in each
region is the first step in variance reduction
and is further enhanced by the use of the neural network
to reduce the required runs of the full calculation itself.
We have demonstrated this with a series of examples
with different degrees of difficulty in different number of dimensions.
We show that this approach can potentially bring great improvement,
most notably to expensive and time consuming calculations
where training a network to preprocess large amounts of points
before getting their true result may be more convenient.

We finalize commenting on ideas and possibilities that can be
contemplated for future extensions.
To keep the discussion on point,
we did not delve into efficient generation of points in the domain space,
however, this work could extended by the addition
of an random space generator that can learn and efficiently generate points in all regions.
For example, a \texttt{vegas} map could be used to accelerate
the generation of points in small regions.
Staying on the ML side,
one could also employ generative neural networks
such as normalizing flows.
Furthermore,
while we show that our way to decide divisions may be advantageous
there is no reason to conclude that it is the best.
As we have demonstrated,
a neural network can be used to apply any type of division,
so now we can extend the question,
could it be possible to use ML
to discover ways to apply stratified sampling
that are closer to the optimal one?

\section{Acknowledgements}

We want to thank Pyungwon Ko, Minho Son and Ahmed Hammad
for useful comments and discussions.
KB is supported by the KIAS Individual Grant. No. PG097601.
The work of RR is supported by a KIAS Individual Grant (QP094601)
via the Quantum Universe Center at Korea Institute for Advanced Study.
The earlier part of this work is supported by National Research Foundation
of Korea NRF-2021R1A2C4002551 (MP).
This work is supported by the Center for Advanced Computation at Korea Institute for Advanced Study.

\bibliographystyle{JHEP}
\bibliography{reference}

\end{document}